\documentclass[twocolumn,superscriptaddress,longbibliography,prc]{revtex4-2}
\usepackage{times}
\usepackage[T1]{fontenc}

\usepackage{physics}
\usepackage{amsmath,amssymb,mathrsfs}
\usepackage{graphicx}
\usepackage{dcolumn}
\usepackage{bm}
\usepackage[subrefformat=parens]{subcaption}
\usepackage{morefloats}
\usepackage{multirow}
\usepackage{amssymb}
\usepackage{amsmath}
\usepackage{xcolor}
\usepackage{fix-cm}
\usepackage{verbatim}
\usepackage{float}
\usepackage{mathptmx} 
\usepackage[T1]{fontenc}
\usepackage[colorlinks,allcolors=blue]{hyperref}

\makeatletter
\def\NAT@def@citea{\def\@citea{\NAT@separator}}
\makeatother

\def\0\\{\nonumber\\}

\def\zI{\mathrm{i}\hspace{0.2mm}}

\newcommand{\beq}{\begin{equation}}
\newcommand{\eeq}{\end{equation}}
\newcommand{\beqn}{\begin{eqnarray}}
\newcommand{\eeqn}{\end{eqnarray}}

\newcommand{\SO}{\mathrm{SO}}
\newcommand{\U}{\mathrm{U}}

\makeatletter
\newcommand\footnoteref[1]{\protected@xdef\@thefnmark{\ref{#1}}\@footnotemark}
\makeatother

\begin{document}

\title{
Formation of bound composite vortices of a singly-quantized $^1$S$_0$ vortex and half-quantized $^3$P$_2$ vortices in the $^1$S$_0$--$^3$P$_2$ coexisting phase in neutron stars
}

\author{Tatsuhiro Hattori}
\email{hattori.t.a850@m.isct.ac.jp}
\affiliation{Department of Physics, School of Science, Institute of Science Tokyo, Tokyo 152-8550, Japan}

\author{Muneto Nitta}
\email{mune.nitta@gmail.com}
\affiliation{Department of Physics $\&$ Research and Education Center for Natural Sciences, Keio University, 4-1-1 Hiyoshi, Yokohama, Kanagawa 223-8521, Japan}
\affiliation{International Institute for Sustainability with Knotted Chiral Meta Matter (WPI-SKCM$^2$), Hiroshima University, 1-3-1 Kagamiyama, Higashi-Hiroshima, Hiroshima 739-8531, Japan}

\author{Kazuyuki Sekizawa}
\email{sekizawa@phys.sci.isct.ac.jp}
\affiliation{Department of Physics, School of Science, Institute of Science Tokyo, Tokyo 152-8550, Japan}
\affiliation{Nuclear Physics Division, Center for Computational Sciences, University of Tsukuba, Ibaraki 305-8577, Japan}
\affiliation{RIKEN Nishina Center, Saitama 351-0198, Japan}

\date{\today}

\begin{abstract}
\edef\oldrightskip{\the\rightskip}
\begin{description}
\rightskip\oldrightskip\relax
\setlength{\parskip}{0pt} 
\item[Background]
Pulsar glitches, sudden spin-ups of rotational frequency of rotating
neutron stars, have been considered to be caused by dynamics of
quantized vortices of superfluid neutrons, especially $^1\text{S}_0$
superfluid in the inner crust of neutron stars. 
In the outer core of neutron stars, neutrons are in the $^3\text{P}_2$
superfluid phase, which can form half-quantized vortices (HQVs), manifesting
distinct characters as compared to usual singly-quantized vortices (SQVs) of
$^1\text{S}_0$ superfluid. It has recently been suggested that the coupling
between these two vortex species near the crust--core boundary gives rise
to a large-scale vortex network, providing a possible explanation of
the observed glitch-size distribution.

\item[Purpose]
We aim to elucidate the effect of interaction between a SQV in
$^1\text{S}_0$ neutron superfluid and HQVs in $^3\text{P}_2$
neutron superfluid, which may be important near the crust--core
boundary in neutron stars, where those two phases may coexist.

\item[Methods]
Using the Gross-Pitaevskii equations (GPE) for the $^1\text{S}_0$
Bose-Einstein condensate (BEC) and $^3\text{P}_2$ spinor BEC,
we perform two-dimensional (2D) simulations of a single $^1\text{S}_0$
SQV and two $^3\text{P}_2$ HQVs in a $^1\text{S}_0$--$^3\text{P}_2$
coexisting phase, varying coupling constants of interaction between
the two species.

\item[Results]
We perform 2D GPE calculations for the $^1\text{S}_0$--$^3\text{P}_2$ coexisting
phase with varying the magnitude of the coupling constants of the
density-density as well as Josephson coupling terms. From the results,
we find that the Josephson coupling, arising from the relative phase between 
the $^1\text{S}_0$ and $^3\text{P}_2$ condensates, induces a strong
attractive interaction between two HQVs and a single SQV, which dominates
over the density-density coupling term. When pinning potentials are applied
to the $^3\text{P}_2$ HQVs and the $^1\text{S}_0$ SQV 
at spatially separated locations, the Josephson-term-driven attraction is
found to be sufficiently strong to overcome the pinning potential,
demonstrating that the inter-condensate Josephson coupling can
significantly alter the spatial configuration of quantized vortices
near the $^1\text{S}_0$--$^3\text{P}_2$ phase boundary.

\item[Conclusions]
The results demonstrate that the Josephson coupling plays a crucial role
in the interaction between $^3\text{P}_2$ HQVs and a $^1\text{S}_0$ SQV
and suggest that two HQVs and a single SQV can form a bound composite
vortex at the phase boundary near the inner crust and the outer core
of neutron stars.

\end{description}
\end{abstract}
\maketitle

\section{Introduction}\label{Sec:Intro}

Neutron stars are among the densest objects in the universe, with central
densities exceeding nuclear saturation density~\cite{Lattimer2016}, making
them a unique laboratory at the intersection of particle physics, nuclear
physics, and astrophysics. Formed as compact remnants of core-collapse
supernovae, neutron stars rotate rapidly, with spin periods ranging from 
approximately $1$\,ms to $10$\,s~\cite{Haskell2018}. A subset of neutron
stars emit a beam of electromagnetic radiation from their magnetic pole.
Because this emission is observed on the Earth as periodic pulses synchronized
with the rotation, these objects are called \textit{pulsars}~\cite{Lattimer2016}.
The interior of neutron stars reaches densities far exceeding those achievable
in terrestrial nuclear experiments, making it the only known realization of
supranuclear-density matter in the universe~\cite{Lattimer2016, Ozel2016}.

Among the observable properties of pulsars, the spin period can be measured
with exceptional precision via radio or X-ray timing~\cite{Haskell2018}.
The neutron star mass can be measured relatively well, especially for binary
systems. In contrast, the radius has remained difficult to constrain and
has only recently been measured to meaningful precision, primarily through
X-ray pulse-profile modeling by the Neutron Star Interior Composition Explorer
(NICER)~\cite{Miller2021, Riley2021}. The magnetic field strength and its
internal configuration remain subject to considerable observational uncertainty.
Thermal X-ray observations of neutron star surfaces provide an additional
window into the stellar interior. For instance, the rapid cooling of the
Cassiopeia~A neutron star, observed in real time by the \textit{Chandra}
X-ray Observatory, has been interpreted as possible direct evidence for
the onset of $^3\text{P}_2$ neutron superfluidity in the stellar
core~\cite{Page2011, Shternin2011,NamSekizawa2026} or any other processes
that cause rapid cooling such as Direct Urca processes \cite{Negreiros2013}.
This could be the first observational indication that the neutron
$^3\text{P}_2$ superfluidity occurs in the outer core of neutron stars
(see, \textit{e.g.}, Refs.~\cite{Graber:2016imq,Sedrakian:2018ydt,Haskell2018,Sedrakian:2026jkj}
for reviews for superfluidity). Furthermore, the frictional motion of
pinned superfluid vortices during spin-down---known as vortex creep---has
been proposed as an internal heating mechanism that may explain the
unexpectedly high surface temperatures of old neutron 
stars~\cite{Alpar1984, Fujiwara_2024, NamSekizawa2025},
connecting vortex dynamics to neutron star thermal observations.

A hallmark observational phenomenon associated with neutron stars is
the \textit{glitch}~\cite{David_Edward_doi:10.1126/science.162.3861.1481,A_Pulsar_Supernova_Association_Nature},
a sudden, quasi-periodic spin-up of the pulsar rotation frequency
\cite{Antonopoulou:2022rpq,Zhou:2022cyp}. The glitch mechanism is
widely believed to involve dynamics of quantized vortices of superfluid
neutrons. The conventional vortex avalanche model attributes glitches
to the sudden unpinning of $^1\text{S}_0$ superfluid vortices in
the inner crust (see, \text{e.g.}, Refs.~\cite{
Anderson:1975zze,
Cheng1988,
Warszawski2011,
Alpar1984,
Mochizuki1995,
Mochizuki1997,
Mochizuki1999,
haensel2007neutron,
PhysRevLett.109.241103,
PhysRevLett.110.011101,
10.1093/mnras/sts108,
PhysRevC.90.015803,
Antonelli_2023vpd},
and references therein). However, a growing number of observations
have revealed glitch events that cannot be explained solely by this
model~\cite{Ray_2019, Dib_2008}, calling for a new theoretical framework
that can describe the full diversity of observed glitch phenomena.

In the outer core of neutron stars, neutrons are expected be in
the $^3\text{P}_2$ superfluid phase
\cite{
Amundsen:1984qc,Takatsuka:1992ga,sauls_nato,Baldo:1992kzz,Elgaroy:1996hp,Khodel:1998hn,Baldo:1998ca,Khodel:2000qw,Zverev:2003ak,Maurizio:2014qsa,Bogner:2009bt,Srinivas:2016kir,
Tabakin:1968zz,Hoffberg:1970vqj,
10.1143/PTP.44.905,10.1143/PTP.46.114,
10.1143/PTP.47.1062,
Richardson:1972xn}, forming a spin-triplet $p$-wave condensate
that is realized on the Earth in only a handful of systems
such as superfluid $^3$He. Moreover, the outer core allows
coexistence of $^3\text{P}_2$ neutron superfluid and $^1\text{S}_0$
proton superconductor, offering rare environment for a multi-component
condensate system of broad interest. On the theoretical side, significant
progress has been made in the microscopic understanding of the $^3\text{P}_2$
superfluid. The Ginzburg-Landau (GL) framework \cite{Richardson:1972xn,
Sauls:1978lna,Masuda:2015jka,Yasui:2018tcr,Yasui:2019unp} enables us
to study the ground states as well as nonuniform states such as quantized
vortices. The ground state is a nematic phase \cite{Sauls:1978lna}, 
and more precisely, it is in the uniaxial (UN), $\text{D}_2$
biaxial nematic ($\text{D}_2$-BN), or $\text{D}_4$ biaxial nematic
($\text{D}_4$-BN) phases depending on the strength of magnetic field
and temperature \cite{Masuda:2015jka,Yasui:2018tcr,Yasui:2019unp}. 
Quantized vortices in the $^3\text{P}_2$ superfluid have been studied
in the GL framework \cite{Muzikar:1980as,Sauls:1982ie,Masuda:2015jka,
Masuda:2016vak,Leinson2020,Kobayashi:2022moc,Kobayashi:2022dae}.
Internal degrees of freedom of the $^3\text{P}_2$ superfluid allow
for presence of half-quantized vortices (HQVs) in strong magnetic
field in which the system is in the $\text{D}_4$-BN phase \cite{Masuda:2016vak}.
These vortices are also characterized by the non-Abelian fundamental
group and are called non-Abelian HQVs. A singly quantized vortex (SQV)
splits into two HQVs, forming a vortex molecule structure \cite{Kobayashi:2022moc}, 
and the interaction between two or more molecules was studied 
\cite{Kobayashi:2022dae}. Other than vortices, domain walls \cite{Yasui:2019vci} 
and surface defects \cite{Yasui:2019pgb} were also studied. There
are also rich massless and massive bosonic excitations~\cite{Bedaque:2003wj,
Leinson:2011wf,Leinson:2012pn,Leinson:2013si,Bedaque:2012bs,bedaquePRC14,
Bedaque:2013fja,Bedaque:2014zta,Leinson:2009nu,Leinson:2010yf,Leinson:2010pk,
Leinson:2010ru,Leinson:2011jr}, which were proposed to explain
cooling and transport phenomena in neutron stars.

Moreover, recently the $^3\text{P}_2$ superfluid has been studied
microscopically by using the Eilenberger and Bogoliubov-de Gennes
(BdG) equations \cite{Mizushima:2016fbn,Mizushima:2019spl,Mizushima:2021qrz} 
which are applicable near the transition temperature and can describe
short distance behaviors such as vortex cores. The complete phase
diagram under strong magnetic field and temperature as well as
topological superfluidity such as topologically-protected gapless
Majorana fermions on surfaces and vortex cores were shown. Subsequently,
the first microscopic study of quantized vortices in the $^3\text{P}_2$
superfluid was made in the BdG equation \cite{Masaki:2019rsz,
Masaki:2021hmk,Masaki:2023rtn}. In particular, a SQV was found
to host two zero-energy Majorana fermion bound states in its core
\cite{Masaki:2019rsz}. Such a SQV splits into two non-Abelian HQVs
\cite{Masaki:2021hmk}, consistent with the analysis in the GL theory,
and each HQV hosts a single zero-energy Majorana fermion bound state.

Separately, the coexistence of $^1\text{S}_0$ and $^3\text{P}_2$
superfluids were studied within the GL framework \cite{Yasui:2020xqb},
obtaining the phase diagram as a function of temperature and magnetic
field. In Ref.~\cite{Yasui:2020xqb}, it was shown that the coupling
between the condensates for the $^1\text{S}_0$ and $^3\text{P}_2$
superfluids is a form of the \textit{Josephson coupling} in the quadratic order
in terms of both the condensations. Further, it was found that the
two condensates can coexist under weak magnetic fields, and the
parameter region of the $\text{D}_4$-BN phase is enhanced to the
whole the $^3\text{P}_2$ superfluid region. These results indicate
that the internal structure of $^3\text{P}_2$ vortices and the
thermodynamic properties of the coexistence phase are strongly
influenced by the interplay between the two superfluid components.
However, a full microscopic simulation of dynamics of $^3\text{P}_2$
superfluid vortices in the coexistence phase has not been carried out yet.

In the outer core, the $^3\text{P}_2$ superfluid phase hosts vortices
with qualitatively different properties from those in the inner crust,
including the possibility of HQVs~\cite{Kobayashi:2022dae}. Building
on this distinction, it was proposed in Ref.~\cite{Marmorini2024}
that the coupling between outer-core $^3\text{P}_2$ vortices and
inner-crust $^1\text{S}_0$ vortices of superfluid neutrons gives
rise to a large-scale vortex network, which has been advocated to
be a candidate mechanism that explains the pulsar glitch phenomenon.
Namely, the vortex network model provides a scaling law of the glitch
energy that can explain the same law from the observation without
fine tuning parameters. Despite this proposal, a microscopic analysis
of structure and dynamics of $^3\text{P}_2$ vortices, as well as
their interaction with proton superconducting flux tubes in the
outer core, has not yet been carried out~\cite{Drummond2017}.
In particular, if the $^3\text{P}_2$ and $^1\text{S}_0$ superfluids
are overlapped in a certain region of neutron stars, vortices in
these superfluids would meet and interact, forming junctions
called \textit{boojums}. These are key ingredient for a large-scale 
vortex network mentioned above. Josephson currents across the
$^3\text{P}_2$--$^1\text{S}_0$ boundary were also studied \cite{Sedrakian:2024dgk}.
The time-dependent oscillating Josephson current induced by vortex
motion produces electromagnetic radiation whose power far exceeds
Ohmic dissipation, potentially contributing to neutron star heating
during the photon cooling era. We mention here that a one-dimensional
system of coexisting s-wave and p-wave superfluids was discussed
in the context of ultracold atomic gases \cite{Guo:2022xhc}.

In this work, we investigate the microscopic interaction between
$^3\text{P}_2$ and $^1\text{S}_0$ superfluid vortices in a
$^3\text{P}_2$--$^1\text{S}_0$ coexisting phase, by extending
the thermodynamic analysis of Ref.~\cite{Yasui:2020xqb} to include
spatially inhomogeneous vortex configurations. We perform two-dimensional
(2D) Gross-Pitaevskii (GP) calculations, which are valid at zero
temperature, treating coefficients of interactions free parameters,
in contrast to the GL formalism valid near the transition
temperature~\footnote{
While the BdG formalism is valid in the whole temperature region,
it is difficult to study dynamics of vortices. We plan to address
this point in a future work.
}.
Building upon the microscopic vortex structure identified in
Ref.~\cite{Masuda:2015jka}, we evaluate how the coupling between
these two vortex species affects the vortex configurations, with
the aim to provide a quantitative basis for understanding the
diversity of observed glitch phenomena beyond the reach of the
conventional vortex avalanche model that relies only on the
$^1\text{S}_0$ superfluid in the inner crust of neutron stars.

Vortex molecule structure in multi-component condensates has
attracted substantial interest in condensed matter physics.
For instance, two HQVs are connected by a domain wall in
two-component Bose-Einstein condensates (BECs) \cite{Son:2001td,
Kasamatsu:2004tvg,Kasamatsu:2005xiu,Cipriani:2013nya,Tylutki:2016mgy,
Eto:2017rfr,PhysRevA.95.023605,Eto:2019uhe,Kobayashi:2018ezm},
two-gap or two-component superconductors \cite{Babaev:2001hv,
Babaev:2002ck,Babaev:2004rm,Goryo:2007qlm}, and chiral $p$-wave
superconductors \cite{PhysRevB.86.060514,Garaud:2015}. Such
configurations were extended to more complicated vortex molecules
in multi-components systems \cite{Nitta:2010yf,Eto:2012rc,
Eto:2013spa,Nitta:2013eaa,PhysRevA.94.023617,Dantas:2015fka}
and/or arbitrary charges \cite{Garaud:2014laa,Chatterjee:2019jez}.
In this work, we show that a vortex molecule of a form of
HQV--SQV--HQV can be formed near the crust--core boundary,
which consists of a single $^1\text{S}_0$ SQV and two
$^3\text{P}_2$ HQVs which are connected to the SQV by
a domain wall due to the quadratic Josephson coupling.

The article is organized as follows. In Sec.~\ref{Sec:Methods},
we introduce the GP framework used to describe coexisting
$^3\text{P}_2$ and $^1\text{S}_0$ superfluids, including
$^3\text{P}_2$ and $^1\text{S}_0$ interaction. In Sec.~\ref{Sec:Details},
we outline the numerical procedure for various vortex configurations.
In Sec.~\ref{Sec:Results}, we present the results of GP calculations
and discuss how vortex configurations change depending on the coupling
constants for the interaction terms. Finally, a summary and prospect
are given in Sec.~\ref{Sec:Conclusion}.

\section{Methods}\label{Sec:Methods}

\subsection{Total free energy functional}\label{sec-total}

We consider the coexisting phase of $^1\text{S}_0$ and $^3\text{P}_2$
neutron superfluids, which may be realized near the crust-core boundary
in neutron stars. To describe such a coexisting phase of $^1\text{S}_0$
and $^3\text{P}_2$ neutron superfluids, we employ the GP equations.
In this article, we denote the complex scalar order parameter of
the $^1\text{S}_0$ superfluid as $\sigma(\mathbf{r})$, while the
five-component complex spinor order parameter of the $^3\text{P}_2$
superfluid is denoted as $\vec{\psi}(\mathbf{r})=(\psi_{-2}(\mathbf{r}),
\psi_{-1}(\mathbf{r}),\psi_0(\mathbf{r}),\psi_1(\mathbf{r}),\psi_2(\mathbf{r}))^\text{T}$.
We restrict our numerical analyses to 2D coordinate space, while the
equations will be given in a general form that can be applied to
three-dimensional (3D) systems.

The total free energy functional of the mixed system of $^1\text{S}_0$
and $^3\text{P}_2$ superfluids is given by
\begin{align}
    F_{\mathrm{total}}[\sigma, \vec{\psi}] 
    = F_{\mathrm{singlet}}[\sigma] 
    + F_{\mathrm{triplet}}[\vec{\psi}] 
    + F_{\mathrm{int}}[\sigma, \vec{\psi}],
    \label{eq:F-total}
\end{align}
where $F_{\mathrm{singlet}}[\sigma]$ and $F_{\mathrm{triplet}}[\vec{\psi}]$ 
are the free energy functionals of the $^1\text{S}_0$ and $^3\text{P}_2$
condensates, respectively, and $F_{\mathrm{int}}[\sigma, \vec{\psi}]$
accounts for contributions from the inter-condensate interaction.
Each term is discussed in detail in the following subsections.
The equilibrium configurations of $\sigma(\mathbf{r})$ and
$\vec{\psi}(\mathbf{r})$ are obtained by minimizing $F_{\mathrm{total}}
[\sigma, \vec{\psi}]$ with respect to the order parameters, yielding
the coupled Euler-Lagrange equations:
\begin{align}
    \frac{\delta F_{\mathrm{total}}}{\delta \sigma^*} &= 0,
    \label{eq:EL-sigma} \\
    \frac{\delta F_{\mathrm{total}}}{\delta \psi_m^*} &= 0 \quad (m = -2, \ldots, 2).
    \label{eq:EL-psi}
\end{align}
In practice, we solve these equations numerically using the
imaginary-time method, as described in Sec.~\ref{Sec:method_im_time}.

\subsection{Free energy functional of the $^1\text{S}_0$ neutron superfluid}\label{sec-1S0}

The $^1\text{S}_0$ neutron superfluid is characterized by a complex
scalar order parameter $\sigma(\mathbf{r})$, which represents the
condensate of spin-singlet Cooper pairs. Within the GP framework,
the free energy functional is given by
\begin{equation}
    F_{\mathrm{signlet}}[\sigma]=\int\left[\frac{\hbar^2}{2M}\bigl|\nabla\sigma(\mathbf{r})\bigr|^2+U_{\mathrm{trap}}(\mathbf{r})\bigl|\sigma(\mathbf{r})\bigr|^2+\frac{g_0}{2}\bigl|\sigma(\mathbf{r})\bigr|^4
    \right]\dd\mathbf{r},
    \label{GPE-1S0}
\end{equation}
where $M = 2m_n$ is the mass of a neutron Cooper pair, being $m_n$
the neutron mass, and $g_0$ is a constant that controls the Landau
potential. The first term is the gradient energy, which penalizes
spatial variations of the order parameter and governs the coherence
length of the condensate. The second term is the trapping potential
that confines the condensates in a finite region. The third term
describes the interaction that governs non-linear effects.

\subsection{Free energy functional of the $^3\text{P}_2$ neutron superfluid}\label{sec-3P2}

Since the total angular momentum of a $^3\text{P}_2$ Cooper pair
is $J = 2$, the $^3\text{P}_2$ neutron superfluid can be formally
modeled as a spin-2 spinor BEC. We therefore adopt the spin-2 GP
framework for spinor BEC~\cite{Kawaguchi2012, Ueda2002} to describe
the $^3\text{P}_2$ condensate. Among possible phases of the spin-2
spinor BECs, the relevant one for the $^3\text{P}_2$ neutron superfluid
is the nematic phase \cite{Song:2007ca,Uchino:2010pf}. The free energy
functional of the $^3\text{P}_2$ superfluid reads~\cite{Kawaguchi2012}
\begin{align}
    F_{\mathrm{triplet}}[\vec{\psi}] = \int &\,\Bigg\{
    \sum^{2}_{m=-2} \psi_m^*(\mathbf{r})
    \bigg[
    -\frac{\hbar^2}{2M}\nabla^2
    + U_{\mathrm{trap}}(\mathbf{r})
    \bigg]\psi_m(\mathbf{r}) \nonumber \\
    &+ \frac{c_0}{2} n^2(\mathbf{r})
    + \frac{c_1}{2}\bigl|\mathbf{F}(\mathbf{r})\bigr|^2
    + \frac{c_2}{10}\bigl|\Psi_{20}(\mathbf{r})\bigr|^2
    \Bigg\}\dd\mathbf{r},
    \label{eq:F-triplet}
\end{align}
where $n(\mathbf{r})$ is the superfluid density of the $^3\text{P}_2$
superfluid, $n(\mathbf{r})=\sum_{m=-2}^{2}|\psi_m(\mathbf{r})|^2$,
$\mathbf{F}(\mathbf{r})$ is the spin-density vector, and $\Psi_{20}
(\mathbf{r})$ is a gauge-invariant scalar quantity characterizing
the condensate. $c_i$ ($i=\{0,1,2\}$) are coupling constants that
characterize the density-density, spin-spin, and spin-singlet pair
interactions, respectively.

Each component of the spin-density vector
$\mathbf{F}(\mathbf{r})$ is defined by
\begin{align}
    F_\nu(\mathbf{r}) = \sum_{m,m'=-2}^{2} \psi_m^*(\mathbf{r})(f_\nu)_{mm'}\psi_{m'}(\mathbf{r}),
\end{align}
where $f_\nu$ ($\nu=\{x,y,z\}$) are the standard $5\times5$ spin
matrices for spin-2, whose explicit forms can be found in, \textit{e.g.},
Ref.~\cite{Kawaguchi2012}:
\begin{align}
    f_x =& \begin{pmatrix}
        0&1&0&0&0\\
        1&0&\sqrt{\frac{3}{2}}&0&0\\
        0&\sqrt{\frac{3}{2}}&0&\sqrt{\frac{3}{2}}&0\\
        0&0&\sqrt{\frac{3}{2}}&0&1\\
        0&0&0&1&0
    \end{pmatrix},\\[2mm]
    f_y =& \begin{pmatrix}
        0&-i&0&0&0\\
        i&0&-\sqrt{\frac{3}{2}}i&0&0\\
        0&\sqrt{\frac{3}{2}}i&0&-\sqrt{\frac{3}{2}}i&0\\
        0&0&\sqrt{\frac{3}{2}}i&0&-1\\
        0&0&0&i&0
    \end{pmatrix},\\[2mm]
    f_z =& \begin{pmatrix}
        2&0&0&0&0\\
        0&1&0&0&0\\
        0&0&0&0&0\\
        0&0&0&-1&0\\
        0&0&0&0&-2
    \end{pmatrix}.
\end{align}

The $^3\text{P}_2$ superfluid order parameter is often represented
using a $3\times3$ complex symmetric matrix $A(\mathbf{r})$, having matrix elements
$A_{ij}(\mathbf{r})$ ($i,j = \{x,y,z\}$), which encodes the spin-triplet, $p$-wave
pairing structure of the condensate. The matrix element $A_{ij}(\mathbf{r})$ can be
decomposed in terms of the magnetic quantum number components $\psi_m(\mathbf{r})$
($m = -2,\ldots,2$) as \cite{Kobayashi2021}
\begin{align}
    A_{ij}(\mathbf{r}) = \sum_{m=-2}^{2} \psi_m(\mathbf{r})
    \left( \mathcal{Y}^{(2)}_m\right)_{ij},
    \label{eq:Aij}
\end{align}
where $\mathcal{Y}^{(2)}_m$ are the rank-2 symmetric
traceless tensors constructed from the spin-2 spherical harmonics,
forming a basis for the irreducible representation of the rotation
group $\SO(3)$ in the spin-2 sector~\cite{Kobayashi2021}.

The gauge-invariant scalar quantity characterizing the condensate,
$\Psi_{20}(\mathbf{r})$, is constructed from the trace of $A^2(\mathbf{r})$
which is invariant under $\SO(3)$ spin rotations, but transforms non-trivially 
under the $\U(1)$ gauge transformation $A_{ij} \to e^{2i\alpha} A_{ij}$,
acquiring a phase $e^{4i\alpha}$. Using the orthogonality of the basis
tensors $\mathcal{Y}^{(2)}_m$, one obtains
\begin{align}
    \Psi_{20}(\mathbf{r}) \equiv \mathrm{tr}\bigl[A^2(\mathbf{r})\bigr]
    = \sum_{m=-2}^{2} (-1)^m \psi_m(\mathbf{r}) \psi_{-m}(\mathbf{r}),
    \label{eq:Psi20-expand}
\end{align}
where the factor $(-1)^m$ arises from the contraction of the basis
tensors, reflecting the time-reversal properties of the spin-2
representation~\cite{Kobayashi2021}.

The quantity $\Psi_{20}$ plays a central role in identifying the positions
of HQVs in the $^3\text{P}_2$ condensate. Under a $2\pi$ rotation around
a HQV, the order parameter $\psi_m$ acquires a phase of $e^{i\pi} = -1$,
leaving the physical state unchanged due to the discrete $\mathbb{Z}_2$
symmetry of the order parameter manifold~\cite{Kobayashi:2022dae}.
Consequently, $\Psi_{20}$, being quadratic in $\psi_m$, acquires a phase of
$e^{2i\pi} = 1$ under a single circuit around a HQV, whereas it winds by
$e^{4i\pi}$ around an SQV. This distinction makes $\Psi_{20}$ a sensitive
probe of the vortex topology: the phase singularities of
$\Psi_{20}(\mathbf{r})$---points around which its phase winds by
$2\pi$---correspond to the cores of individual HQVs, as opposed to
the integer vortices of $\sigma(\mathbf{r})$. The vortex core positions
of the HQVs are therefore identified numerically by locating the phase
singularities of $\Psi_{20}(\mathbf{r})$ on the simulation lattice.

\subsection{Free energy functional of inter-condensate interaction}\label{sec-interaction}

In the coexistence region near the crust-core boundary, both the
$^1\text{S}_0$ and $^3\text{P}_2$ superfluid phases may be present
simultaneously. In such a $^1\text{S}_0$--$^3\text{P}_2$ coexisting
region, the two order parameters $\vec{\psi}(\mathbf{r})$ and $\sigma(\mathbf{r})$
are coupled through an interaction term in the free energy functional.
Following the GL framework, and adopting the structure of the coupling term
introduced in Ref.~\cite{Yasui:2020xqb}, the interaction free energy
between the two condensates is composed of two terms:
\begin{align}
F_{\mathrm{int}}[\vec{\psi},\sigma]
= \int\Bigl[\mathcal{E}_{\text{den}}(\mathbf{r})+\mathcal{E}_{\text{Jos}}(\mathbf{r})\Bigr]\dd\mathbf{r}
    \label{eq:Fint}
\end{align}
where $\mathcal{E}_\text{den}(\mathbf{r})$ and $\mathcal{E}_\text{Jos}(\mathbf{r})$
represent the interaction energy densities associated with the density-density
and the Josephson couplings, respectively. They are given by
\begin{align}
    \mathcal{E}_{\mathrm{den}}(\mathbf{r})=&\zeta_1|\sigma(\mathbf{r})|^2\mathrm{tr}[A^*(\mathbf{r})A(\mathbf{r})]\nonumber\\
    =&\zeta_1|\sigma(\mathbf{r})|^2\left[\sum^2_{m=-2}(-1)^m|\psi_m(\mathbf{r})|^2\right]
    ,\label{eq:Eden_den}\\[2mm]
    \mathcal{E}_{\mathrm{Jos}}(\mathbf{r}) =&\zeta_2(\sigma^2\mathrm{tr}[A^{*2}(\mathbf{r})]+\sigma^{*2}\mathrm{tr}[A^2(\mathbf{r})]) 
    \nonumber\\
    =& \zeta_2 \sum^2_{m=-2} (-1)^m\left[
    \sigma^2(\mathbf{r})  \psi_m^{*2}(\mathbf{r}) 
    + \sigma^{*2}(\mathbf{r})  \psi_m^{2}(\mathbf{r}) 
    \right]\nonumber\\=&\zeta_2\sum^2_{m=-2}(-1)^m|\sigma(\mathbf{r}) |^2|\psi_m(\mathbf{r}) |^2\cos[2\Delta\theta_m(\mathbf{r})],
    \label{eq:Eden_Jos}
\end{align}
where $\Delta\theta_m(\mathbf{r})\equiv \theta_\sigma(\mathbf{r})
-\theta_m(\mathbf{r})$ denotes the phase difference between the
$^1\text{S}_0$ and $^3\text{P}_2$ order parameters, where
$\theta_\sigma(\mathbf{r})=\mathrm{arg}(\sigma(\mathbf{r}))$ and
$\theta_m(\mathbf{r})=\mathrm{arg}(\psi_m(\mathbf{r}))$ denote
their phase fields, respectively. $\zeta_1$ and $\zeta_2$ are
real coupling constants defined as
\begin{equation}
    \zeta_{1,2} = \tilde{\zeta}_{1,2}\frac{4\pi\hbar^2}{M},
    \label{eq:zeta_dim}
\end{equation}
with parameters $\tilde{\zeta}_{1,2}$.
This normalization is adopted so that $\zeta_{1,2}$ carry the same dimensions 
as the coupling constant $g_0$ in Eqs.~\eqref{GPE-1S0} and \eqref{eq:F-triplet},
allowing for a direct comparison of the interaction strength between the two
condensates. The dimensionless values $\tilde{\zeta}_{1,2}$ are varied
independently  in the present numerical analysis.

The total interaction energy associated with the density-density coupling,
$E_\text{den}=\int\mathcal{E}_\text{den}(\mathbf{r})\dd\mathbf{r}$, is invariant
under independent $\U(1)$ phase rotations of each order parameter. The total
interaction energy associated with the Josephson coupling,
$E_\text{Jos}=\int\mathcal{E}_\text{Jos}(\mathbf{r})\dd\mathbf{r}$, 
represents a coherent pair-exchange interaction between the two superfluid
components, which breaks the relative $\U(1)$ phase symmetry between
$\vec{\psi}(\mathbf{r})$ and $\sigma(\mathbf{r})$. This type of coupling
is analogous to the Josephson interaction known in coupled
superconductors~\cite{JOSEPHSON1962251,Chatterjee:2019jez}, and plays
a crucial role in determining the structure and dynamics of quantized
vortices near the phase boundary between the inner crust and the outer core.
This coupling is analogous to the Rabi coupling in the context of BECs,
in which it is linear with respect to both the condensates. In this work,
we refer to this coupling as the Josephson coupling and associated energy
$E_\text{Jos}$ as Josephson energy. The relative magnitude and signs
of $\zeta_1$ and $\zeta_2$ control whether the interaction is attractive
or repulsive between the two condensates, and are determined from
microscopic considerations in the present analysis.

\section{Computational Details}\label{Sec:Details}

\subsection{Trapping potential and discretization}\label{Sec:method_trap_discretize}

Numerical simulations are performed on a 2D square lattice,
where all fields are represented on a uniform mesh. In this work,
we use $100\times100$ grid points with a $0.1$-fm mesh. To compute
Laplacian of spatial functions, we employ the spectral method executing
fast Fourier transform (FFT) routines of the \texttt{FFTW} library.
Because of the use of FFT, the periodic boundary conditions are
automatically adopted. To avoid vortices to interact due to the
periodicity of the computational box, we confine the condensate
within a trapping potential $U_{\mathrm{trap}}(\mathbf{r})$, which
is defined as a function  of the radial distance $r = |\mathbf{r}|$
from the center of the simulation box. The potential is characterized
by a trap radius $R_{\mathrm{trap}}$ and a smoothing interval $\delta$,
and is given by
\begin{align}
    &U_{\mathrm{trap}}(r) = \nonumber\\
    &\begin{cases}
        0 & r < R_{\mathrm{trap}} - \delta, \\[6pt]
        \displaystyle V \left(\frac{1}{2} - \frac{1}{2}
        \tanh\!\left[\pi\tan\!\left(
        \frac{\pi}{2\delta}(r - R_{\mathrm{trap}})\right)\right]\right)^{-1}
        & R_{\mathrm{trap}} - \delta \leq r\\& \leq R_{\mathrm{trap}} + \delta, \\[6pt]
        V & r > R_{\mathrm{trap}} + \delta,
    \end{cases}
    \label{eq:Utrap}
\end{align}
where $V$ is a large positive constant that effectively excludes the condensate from the region outside the trap.
The intermediate region, $R_{\mathrm{trap}} - \delta \leq r \leq R_{\mathrm{trap}} + \delta$,
connects smoothly the flat interior and the hard wall,
suppressing numerical artifacts arising from an abrupt potential change.
In the present work, we use $V=10^4$\,MeV, $R_{\mathrm{trap}}=4.5$\,fm,
and $\delta=0.4$\,fm for all the simulations given below.

\subsection{Pinning potential}\label{Sec:method_pin}

To independently control the positions of the $^1\text{S}_0$ SQV
and the $^3\text{P}_2$ HQVs, we introduce pinning potentials for
each condensate separately. For the $^1\text{S}_0$ condensate,
we introduce a pinning potential as
\begin{align}
    F_{\mathrm{pin,singlet}} = \int \, 
    U_{\mathrm{pin,singlet}}(\mathbf{r})\, |\sigma(\mathbf{r})|^2\dd\mathbf{r},
    \label{eq:pin-singlet}
\end{align}
which pins the SQV core of the $^1\text{S}_0$ condensate at a separate location.
For the $^3\text{P}_2$ condensate, a pinning potential is coupled to the gauge-invariant 
density $|\Psi_{20}(\mathbf{r})|^2$, adding the following term to the free energy functional:
\begin{align}
    F_{\mathrm{pin.triplet}} = \int \, 
    U_{\mathrm{pin,triplet}}(\mathbf{r})\, |\Psi_{20}(\mathbf{r})|^2\dd\mathbf{r}.
    \label{eq:pin-triplet}
\end{align}
Since $\Psi_{20}$ vanishes at the core of each HQV [see Eq.~\eqref{eq:Psi20-expand}],
this term energetically favors the suppression of $|\Psi_{20}|^2$ at the potential center,
thereby pinning the HQV cores at prescribed locations. The GP equations are supplemented
with the following terms associated with the pinning potentials:
\begin{align}
    \frac{\delta F_{\mathrm{pin,singlet}}}{\delta \sigma^*}=& U_{\mathrm{pin,singlet}}(\mathbf{r})\sigma(\mathbf{r}),\label{eq:derivative_pin_signlet}\\
     \frac{\delta F_{\mathrm{pin,triplet}}}{\delta \psi_m^*}=& (-1)^mU_{\mathrm{pin,triplet}}(\mathbf{r})\psi_{-m}^*(\mathbf{r})\Psi_{\mathrm{20}}(\mathbf{r}).\label{eq:derivative_pin_triplet}  
\end{align}
For the pinning potential, we use a simple form defined as
\begin{align}
    U_{\mathrm{pin,singlet}}(\mathbf{r}) =& \begin{cases}
        V&(r\leq R_{\mathrm{pin}})\\
        0&(r> R_{\mathrm{pin}})
    \end{cases},\label{eq_pin_singlet}\\[2mm]
    U_{\mathrm{pin,triplet}}(\mathbf{r}) =& \begin{cases}
        V&(|\mathbf{r}\pm d_{\mathrm{pin}}\hat{\mathbf{x}}|\leq R_{\mathrm{pin}})\\
        0&\mathrm{elsewhere}
\end{cases}.\label{eq_pin_triplet}
\end{align} 
That is, each pinning site is modeled as a repulsive potential 
with a small spatial extent, where $R_\text{pin}=0.3$\,fm is 
used in the present study.

\subsection{Vortex initialization}\label{Sec:method_vortex_ini}

To obtain solutions with quantized vortices, we introduce the desired phase
winding to the initial order parameters by hand.
Specifically, a phase profile corresponding to the target vortex number is assigned to 
$\vec{\psi}$ and $\sigma$ as initial conditions, \textit{e.g.}, in the case of
the $^1\text{S}_0$ condensate,
\begin{equation}
    \sigma(\mathbf{r}) = \bigl|\sigma_0(\mathbf{r})\bigr|\exp[\zI N_{\mathrm{v}}\theta_\sigma(\mathbf{r})],
\end{equation}
where $\sigma_0$ is the initial guess of the order parameter and
$\theta_\sigma(\mathbf{r})=\arctan\bigl[(y-y_0)/(x-x_0)\bigr]$ is its phase,
rotating about the center of the box $(x_0,y_0)$. In the analyses presented
in this paper, the widing number is always set to $N_{\mathrm{v}}=1$.
We adopt the same procedure also for all the components of the $^3\text{P}_2$
condensate $\psi_m(\mathbf{r})$ ($m=-2,\dots,2$). As the initial guess of the
order parameters, we use a uniform distribution within the trap radius
$R_\text{trap}$. Starting with the initial configuration, the system
is evolved in imaginary time to relax toward the energetically favored
vortex configuration, automatically adjusting the amplitude and the
phase. This procedure yields the equilibrium spatial arrangement and
number of vortices for a given set of parameters.

\subsection{Imaginary-time evolution}\label{Sec:method_im_time}

The GP equations are obtained by taking the functional derivatives
of the total free energy functional, supplemented with the contributions
of the pinning potentials, with respect to the order parameters:
\begin{align}
    i\hbar \frac{\partial \sigma(\mathbf{r})}{\partial t} 
    = \frac{\delta F_{\mathrm{total}}}{\delta \sigma^*(\mathbf{r})},
    \label{eq:TDGPE_1}\\
    i\hbar \frac{\partial \psi_m(\mathbf{r})}{\partial t} 
    = \frac{\delta F_{\mathrm{total}}}{\delta \psi_m^*(\mathbf{r})},
    \label{eq:TDGPE_2}
\end{align}
yielding the coupled equations for $\sigma(\mathbf{r})$ and
$\psi_m(\mathbf{r})$~\cite{Kawaguchi2012}. Equilibrium vortex
configurations are obtained by evolving Eqs.~\eqref{eq:TDGPE_1}
and \eqref{eq:TDGPE_2} in imaginary time ($t \to -i\tau$).

Specifically, in the imaginary time evolution, we solve the
following equations:
\begin{align}
    \nonumber
    -\hbar\frac{\partial\sigma(\mathbf{r})}{\partial\tau}
    &=\left[-\frac{\hbar^2}{2M}\nabla^2+U_{\mathrm{trap}}(\mathbf{r})+g_0|\sigma(\mathbf{r})|^2\right]\sigma(\mathbf{r})\nonumber\\
    &+\sum^2_{m=-2}\left[\zeta_1|\psi_m(\mathbf{r})|^2\sigma(\mathbf{r})+2(-1)^m\zeta_2\sigma^*(\mathbf{r})\psi_m(\mathbf{r})\right]\nonumber\\[1mm]
    &+U_{\mathrm{pin,singlet}}(\mathbf{r})\sigma(\mathbf{r}),\label{eq:im-time-evolv-singlet}\\[2mm]
    -\hbar\frac{\partial\psi_m(\mathbf{r})}{\partial\tau}
    &=\left[-\frac{\hbar^2}{2M}\nabla^2+U_{\mathrm{trap}}(\mathbf{r})+c_0n(\mathbf{r})\right]\psi_m(\mathbf{r})\nonumber\\
    &+\sum^2_{m'=-2}c_1\mathbf{F}^*(\mathbf{r})\mathbf{\cdot}(\mathbf{f})_{mm'}\psi_{m'}(\mathbf{r})+c_2\frac{1}5\psi_m^*(\mathbf{r})\Psi_{20}(\mathbf{r})\nonumber\\[1mm]
    &+\zeta_1|\sigma(\mathbf{r})|^2\psi_m(\mathbf{r})+2(-1)^m\zeta_2\psi_m^*(\mathbf{r})\sigma^2(\mathbf{r})\nonumber\\[2.5mm]
    &+(-1)^mU_{\mathrm{pin,triplet}}(\mathbf{r})\psi_{-m}^*(\mathbf{r})\Psi_{20}(\mathbf{r}),\label{eq:im-time-evolv-triplet}
\end{align}
where $\mathbf{f}=(f_x,f_y,f_z)$ and $\tau$ represents the imaginary time.

At each step of the imaginary-time evolution, the order parameters are
normalized so that the integrated particle numbers of the $^1\text{S}_0$
and $^3\text{P}_2$ condensates are separately conserved. Specifically,
after each step, $\sigma(\mathbf{r})$ and $\vec{\psi}(\mathbf{r})$ 
are normalized such that
\begin{align}
    \int \, |\sigma(\mathbf{r})|^2\dd\mathbf{r} = N, 
    \qquad
    \int \, \sum_{m=-2}^{2} |\psi_m(\mathbf{r})|^2\dd\mathbf{r} = N,
    \label{eq:normalization}
\end{align}
where the particle number $N$ is held fixed and equal for both condensates
throughout the evolution. This normalization condition ensures that the
imaginary-time evolution converges to the ground state at the fixed particle
number for each component, rather than allowing the condensate densities
to change freely.

\subsection{Vortex pinning and activation of the interaction term}
\label{Sec:Pinning_methods}

To analyze the change of vortex configurations triggered by the
$^1\text{S}_0$--$^3\text{P}_2$ coupling, we adopt the following
two-stage protocol. In the first stage, the coupling constants
are set to $\zeta_1 = \zeta_2 = 0$, completely decoupling the
two superfluid components. In this decoupled state, artificial
pinning potentials are introduced for both the singlet and triplet
order parameters independently. The imaginary-time evolution is
performed for the first $100$\,fm/$c$ starting from the initial
state with phase winding at $r=0$. Through the imaginary-time
evolution, the vortices gradually move and are eventually
pinned to the pinning sites.

In the second stage, we switch on the interaction terms $\zeta_1$
and $\zeta_2$ at $\tau=100$\,fm/$c$, activating the coupling between
the $^1\text{S}_0$ and $^3\text{P}_2$ condensates. This protocol
offers a way to numerically obtain the equilibrium configuration
with the $^1\text{S}_0$--$^3\text{P}_2$ couplings, starting from
the well-converged solution for the case without the coupling term.

\subsection{Vortex position detection}

A quantum vortex is characterized by a phase singularity of the order
parameter, near which the density of the order parameter vanishes. 
In the present calculations, the positions of quantum vortices are
extracted by examining the phase of the order parameter.

As an example, the position of a vortex in the $^1\text{S}_0$ order
parameter $\sigma$ is determined as follows. At each lattice point
$(i_x, i_y)$, we collect the phase $\mathrm{arg}(\sigma) =
\mathrm{atan2}(\mathrm{Re}(\sigma), \mathrm{Im}(\sigma))$ 
at the surrounding eight points $\{(i_x-1,i_y-1),\,(i_x,i_y-1),\,
(i_x+1,i_y-1),\,(i_x+1,i_y),\,(i_x+1,i_y+1),\,(i_x,i_y+1),\,(i_x-1,i_y+1),\,
(i_x-1,i_y)\}$. The phase differences $\delta\theta$ between adjacent
points along this loop are then summed. When the absolute value of
a phase difference satisfies $|\delta\theta|>\pi$, the value is
corrected to lie within $[-\pi,\pi]$, accounting for the branch
of $\mathrm{atan2}$ whose range is $(-\pi,\pi]$, so that the phase
is treated as a multivalued function. After applying this correction,
the phase differences at the eight surrounding points are summed.
We confirm that this sum is almost zero if the eight spatial
points do not include the singular point. When the summed value
is nearly $2\pi$, we regard that a vortex is present at the point
$(i_x,i_y)$. The same procedure is applied to the $^3\text{P}_2$
order parameter $\Psi_{20}$. Since $\Psi_{20}\propto(\vec{\psi})^2$,
the total phase winding around a single HQV represented $\Psi_{20}$ is $2\pi$.

Since a single quantum of vorticity $\kappa = 2\pi$ gives rise to two HQVs,
the distance between the two HQVs serves as a useful measure for analyzing 
the interaction between the SQV and the HQVs. This inter-HQV distance
is computed by determining the coordinates of each HQV and evaluating
the Euclidean distance between them. Note that since the vortex detection
is based on the phase winding integral around eight surrounding lattice
points, the spatial resolution of the inter-vortex distance is
fundamentally limited by the lattice spacing of the discretized
computational domain.

\section{Results and Discussion}\label{Sec:Results}

\subsection{Simulations without pinning sites}\label{SubSec:results_nopin}

\begin{figure}[t]\centering
\begin{subfigure}{0.48\textwidth}
    \centering
    \includegraphics[width=0.85\linewidth]{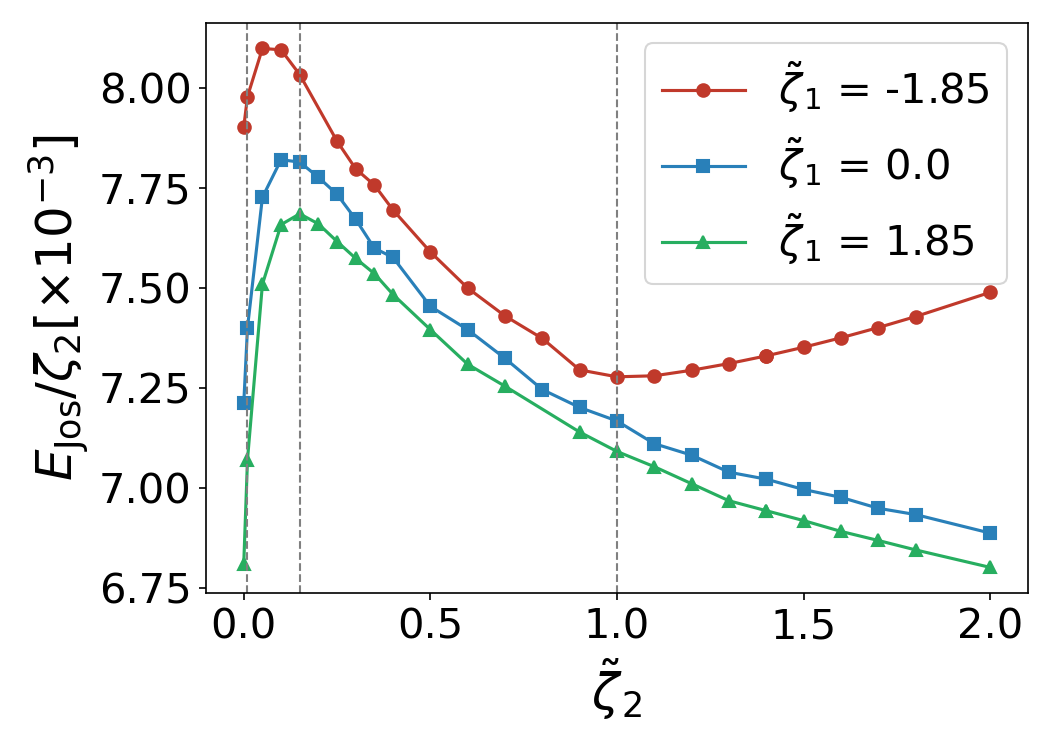}
    \caption{$E_{\mathrm{Jos}}/\zeta_2$}
    \label{fig:Rabi_no_trap_per_zeta2}
\end{subfigure}
\begin{subfigure}{0.48\textwidth}
    \centering
    \includegraphics[width=0.85\linewidth]{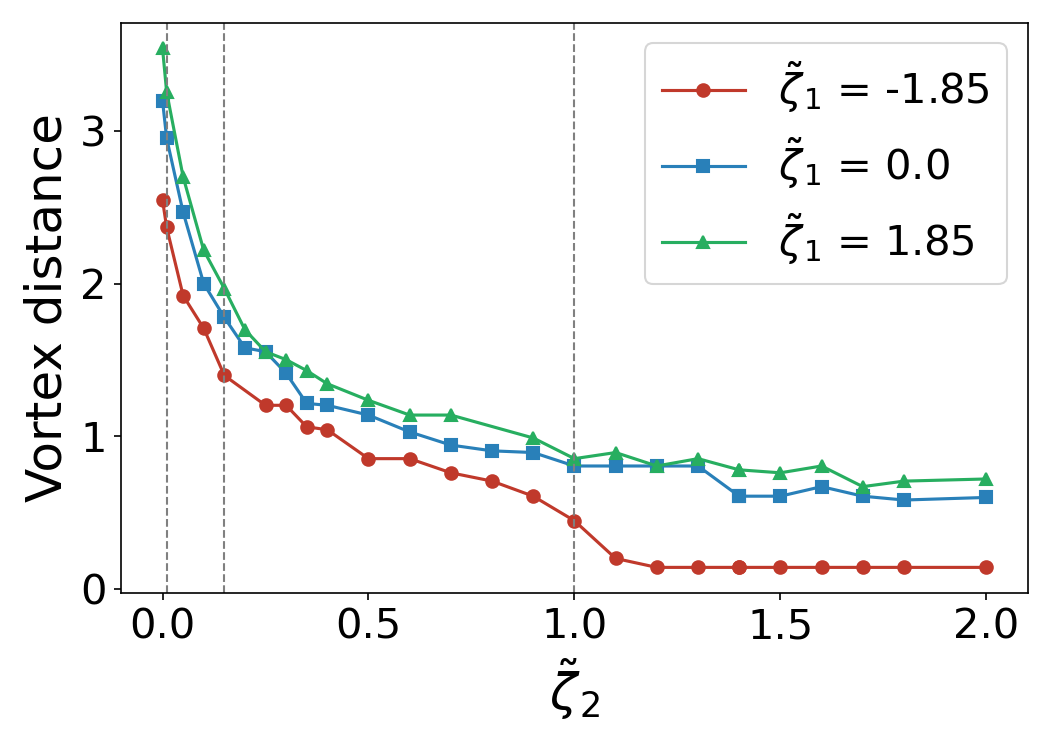}
    \caption{HQVs distance}
    \label{fig:Rabi_no_trap_distance}
\end{subfigure}
\caption{
(a) Josephson energy, $E_{\mathrm{Jos}}/\zeta_2$, as a function 
    of the Josephson coupling constant $\tilde{\zeta}_2$ for three values of 
    the density--density coupling constant $\tilde{\zeta}_1$, as indicated
    in the legend. The system consists of one SQV in the $^1\text{S}_0$
    condensate and two HQVs in the $^3\text{P}_2$ condensate.
(b) Inter-vortex distance between the two HQVs in the $^3\text{P}_2$ 
    condensate as a function of the Josephson coupling constant $\tilde{\zeta}_2$ 
    for three values of the density--density coupling constant $\tilde{\zeta}_1$, 
    as indicated in the legend. The system configuration and imaginary-time
    evolution procedure are the same as in Fig.~\ref{fig:Rabi_no_trap_per_zeta2}.
}\label{fig:no_trap}
\end{figure}

In this section, we analyze how the $^1\text{S}_0$--$^3\text{P}_2$
interaction affects the vortex configuration. To this end, we calculate
the vortex configurations for various $\zeta_1$ and $\zeta_2$ values,
without introducing pinning site.

\subsubsection{Energy and Vortex Distance}\label{subsec:energy_distance_nopin}

In Figs.~\ref{fig:no_trap}(a) and \ref{fig:no_trap}(b), we show
the Josephson energy, $E_{\mathrm{Jos}}/\zeta_2$, and the distance
between two HQVs in the $^3\text{P}_2$ condensate, respectively,
as a function of the coupling constant $\tilde{\zeta}_2$, for three
representative values of $\tilde{\zeta}_1$. Results for $\tilde{\zeta}_1=-1.85$
(attractive density-density coupling), $0$ (no coupling), and $1.85$
(repulsive density-density coupling) are shown by red filled circles,
blue filled squares, and green filled triangles, respectively, connected
with solid lines. Since $E_{\mathrm{Jos}}$ is linear in $\zeta_2$ by
construction [see Eq.~\eqref{eq:Eden_Jos}], plotting $E_{\mathrm{Jos}}/\zeta_2$ 
isolates the geometric contribution arising purely from the phase
structure of the vortices, independently of the coupling strength.

Even for the $\tilde{\zeta}_2 = 0$ case, the Josephson energy shows a small,
but finite value, because $^1\text{S}_0$ and $^3\text{P}_2$ condensate
are independent from each other and, thus, there exists, in general,
finite phase difference among them. When the Josephson coupling is introduced
($0<\tilde{\zeta}_2\lesssim 0.1$), $E_\text{Jos}/\zeta_2$ increases
rapidly, developing characteristic phase structure in the $^1\text{S}_0$
and $^3\text{P}_2$ condensates. As the $\tilde{\zeta}_2$ increases
further, the Josephson energy decreases with decreasing vortex
distance, see Fig.~\ref{fig:no_trap}(b). That is,
since the Josephson coupling costs additional energy associated
with the phase difference, the system tends to reduce the distance
between two $^3\text{P}_2$ HQVs, as a result of energy balance between
gain with reducing $E_\text{Jos}$ and loss associated with the repulsive
interaction between two HQVs. This trend continues up to $\tilde{\zeta}_2=2$
for the $\tilde{\zeta}_1=0$ and $1.85$ cases.

Figure~\ref{fig:Rabi_no_trap_distance} also indicates that the sign of
$\zeta_1$ affects the vortex distance. The positive (negative) $\zeta_1$
increases (decreases) the distance compared with the case of $\zeta_1=0$. 
This implies that positive (negative) $\zeta_1$ yields repulsion (attraction)
between the vortices. This is consistent with the case of two-component BECs \cite{Eto:2011wp}. In particular, when $\zeta_1$ is attractive, for
$\tilde{\zeta}_2\gtrsim 1$, the vortex distance decreases quickly
and eventually reaches a small value of $0.1$--$0.2$\,fm. It is because of
the fact that $\tilde{\zeta}_1<0$ corresponds to attraction for a lager density
overlap, thereby the system prefers to place $^1\text{S}_0$ and $^3\text{P}_2$
vortices as close as possible. We note that since we work with mesh spacing
of $\Delta x=0.1$\,fm, we can not resolve vortex distance shorter than
$\mathcal{O}(\Delta x)$. The minimum value of vortex distance in
Fig.~\ref{fig:Rabi_no_trap_distance} is 0.14\,fm, this value is same
as $\sqrt2\Delta x$. At this stage, both of the phases of $^1\text{S}_0$
and $^3\text{P}_2$ condensates rotate by $2\pi$ around the center of the box,
locking the phase difference at a certain value. As $\zeta_2$ increases, 
the size of the triplet vortex which contains two overlapping HQVs
slightly decreases, which results in an increase of overlap of singlet
and triplet order parameters. As a result, $E_{\mathrm{Jos}}/\zeta_2$
increases with $\tilde{\zeta}_2$ for $1<\tilde{\zeta}_2\le2$, where
the factor of $|\sigma(\mathbf{r})|^2|\psi_m(\mathbf{r})|^2$ in
Eq.~\eqref{eq:Eden_Jos} is enhanced because of the larger density
overlap.

\subsubsection{Vortex Structure}\label{subsec:structure_nopin}

\begin{figure}[tbp]
    \begin{subfigure}[b]{0.48\columnwidth}
        \centering
        \includegraphics[width=\textwidth]{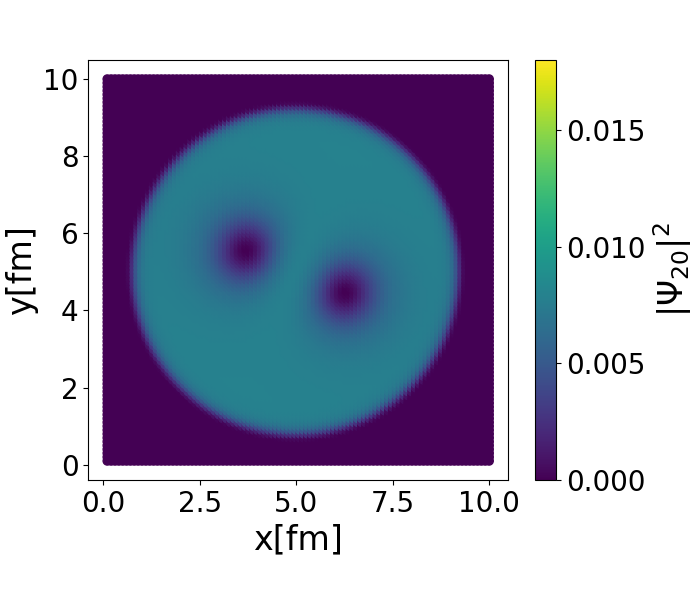}
        \caption{$|\Psi_{20}(\mathbf{r})|^2$}
        \label{fig:nopin_0d0_1d-2_GS_Psi20}
    \end{subfigure}
    \hfill
    \begin{subfigure}[b]{0.48\columnwidth}
        \centering
        \includegraphics[width=\textwidth]{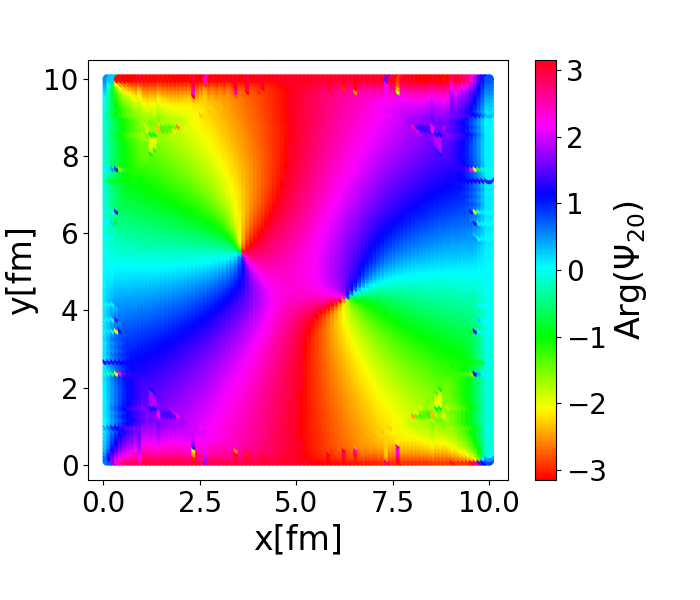}
        \caption{$\arg[\Psi_{20}(\mathbf{r})]$}
        \label{fig:nopin_0d0_1d-2_GS_phase_Psi20}
    \end{subfigure}
    \begin{subfigure}[b]{0.48\columnwidth}
        \centering
        \includegraphics[width=\textwidth]{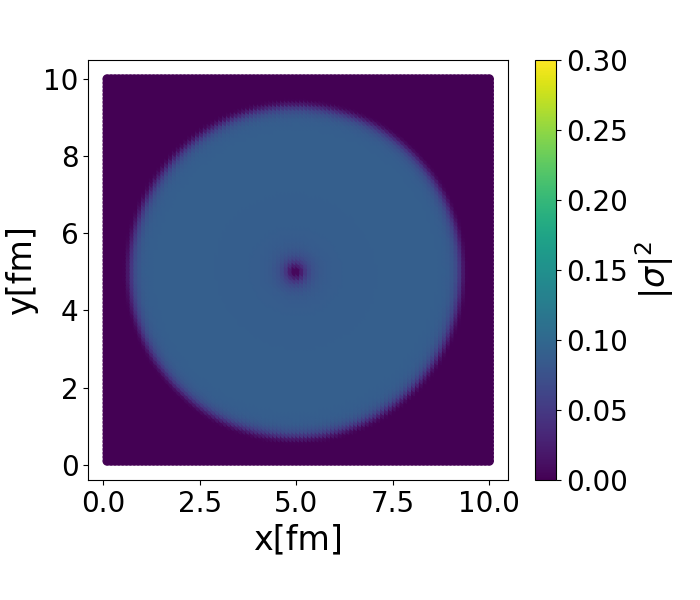}
        \caption{$|\sigma(\mathbf{r})|^2$}
        \label{fig:nopin_0d0_1d-2_GS_rho_s}
    \end{subfigure}
    \hfill
    \begin{subfigure}[b]{0.48\columnwidth}
        \centering
        \includegraphics[width=\textwidth]{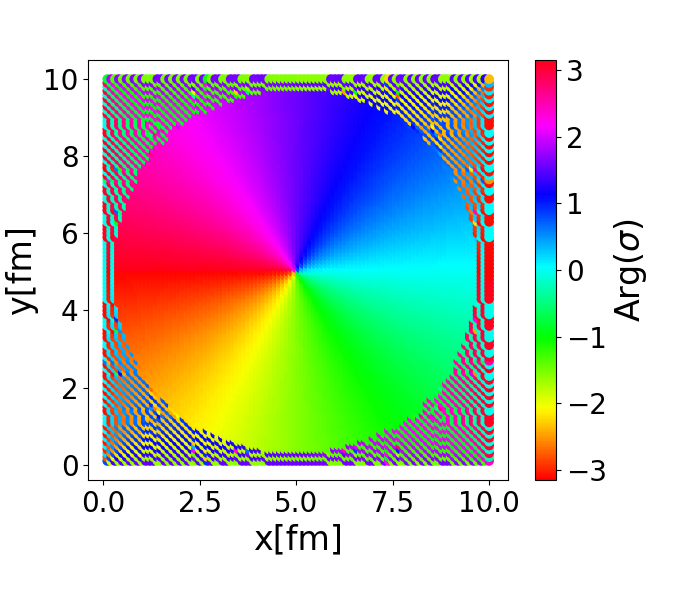}
        \caption{$\arg[\sigma(\mathbf{r})]$}
        \label{fig:nopin_0d0_1d-2_GS_phase_s}
    \end{subfigure}
    \vspace{0.5em}
    \begin{subfigure}[tb]{0.8\columnwidth}
        \centering
        \includegraphics[width=\textwidth]{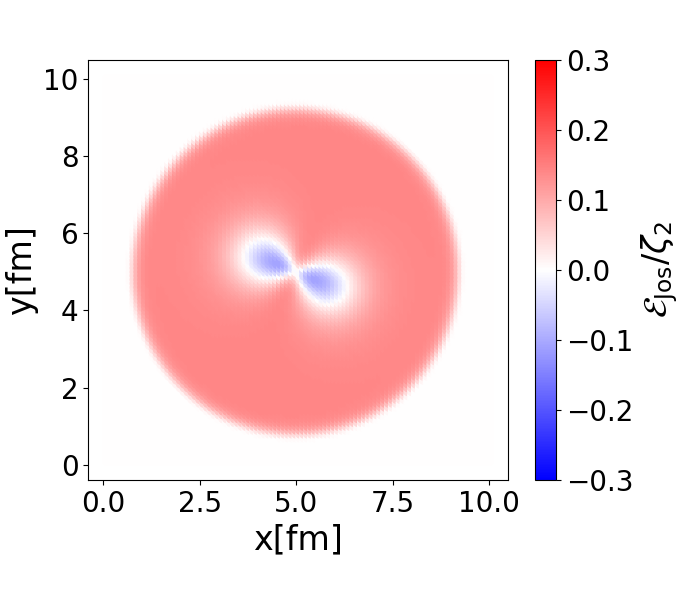}
        \caption{$\mathcal{E}_{\mathrm{Jos}}(\mathbf{r})/\zeta_2$}
        \label{fig:nopin_0d0_1d-2_GS_Rabi}
    \end{subfigure}
    \caption{
        Color maps of the equilibrium configuration obtained by
        imaginary-time evolution without pinning potentials for both
        condensates and $\zeta_1=0,\tilde\zeta_2=0.01$.
        (a) Absolute value of the $^3\text{P}_2$ condensate, $|\Psi_{20}(\mathbf{r})|^2$, 
        showing the two HQV cores as density-depleted regions.
        (b) Phase of the $^3\text{P}_2$ condensate, $\arg[\Psi_{20}(\mathbf{r})]$,
        displaying the $2\pi$ phase winding around each HQV core.
        (c) Absolute value of the $^1\text{S}_0$ condensate, $|\sigma(\mathbf{r})|^2$ 
        showing the SQV core as a single density-depleted region.
        (d) Phase of the $^1\text{S}_0$ condensate, $\arg[\sigma(\mathbf{r})]$
        displaying the $2\pi$ phase winding around the SQV core.
        (e) Spatial distribution of the Josephson energy density,
        $\mathcal{E}_{\text{Jos}}(\mathbf{r})/\zeta_2$,
        reflecting the relative phase between the two condensates.
        All panels share the same spatial coordinates $(x, y)$.
    }
    \label{fig:nopin_all_1d-2}
\end{figure}

\begin{figure}[tbp]
    \begin{subfigure}[b]{0.48\columnwidth}
        \centering
        \includegraphics[width=\textwidth]{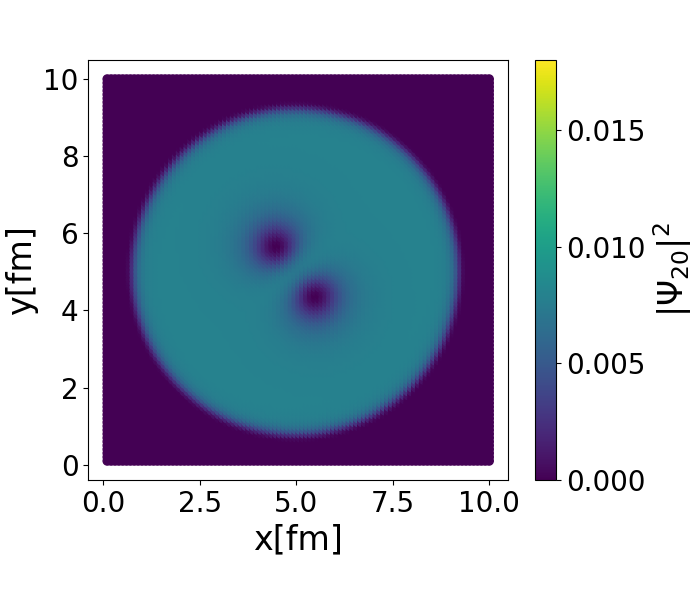}
        \caption{$|\Psi_{20}(\mathbf{r})|^2$}
        \label{fig:nopin_0d0_15d-2_GS_Psi20}
    \end{subfigure}
    \hfill
    \begin{subfigure}[b]{0.48\columnwidth}
        \centering
        \includegraphics[width=\textwidth]{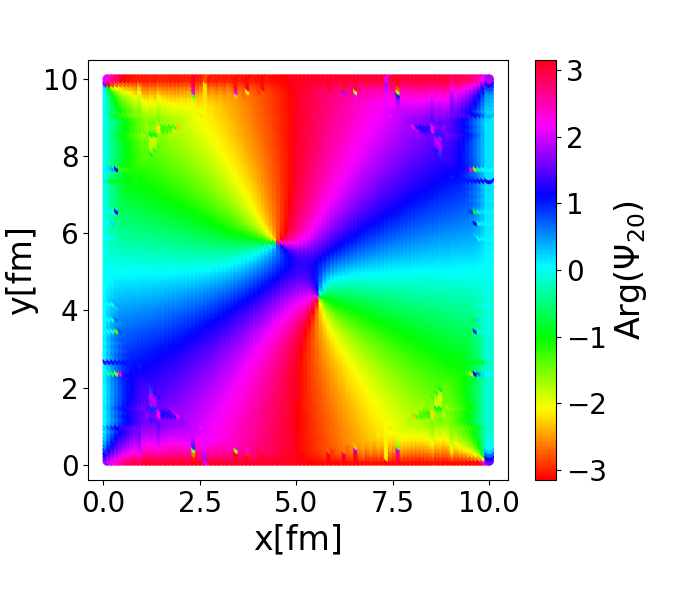}
        \caption{$\arg[\Psi_{20}(\mathbf{r})]$}
        \label{fig:nopin_0d0_15d-2_GS_phase_Psi20}
    \end{subfigure}
    \begin{subfigure}[b]{0.48\columnwidth}
        \centering
        \includegraphics[width=\textwidth]{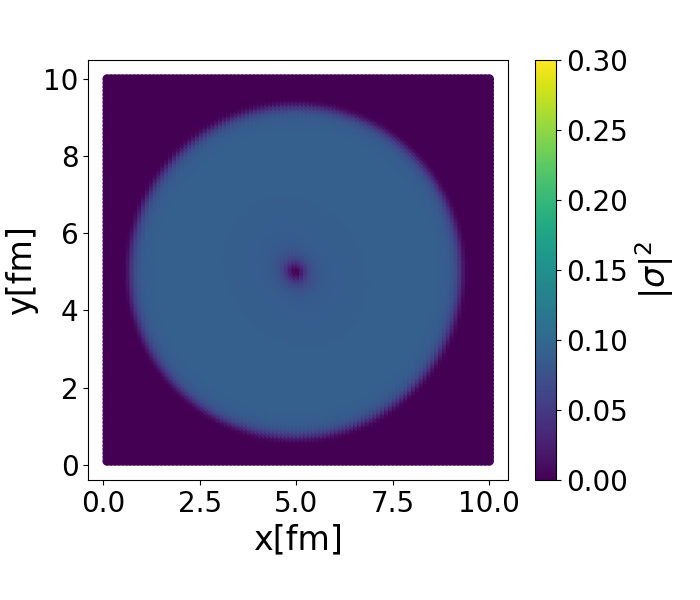}
        \caption{$|\sigma(\mathbf{r})|^2$}
        \label{fig:nopin_0d0_15d-2_GS_rho_s}
    \end{subfigure}
    \hfill
    \begin{subfigure}[b]{0.48\columnwidth}
        \centering
        \includegraphics[width=\textwidth]{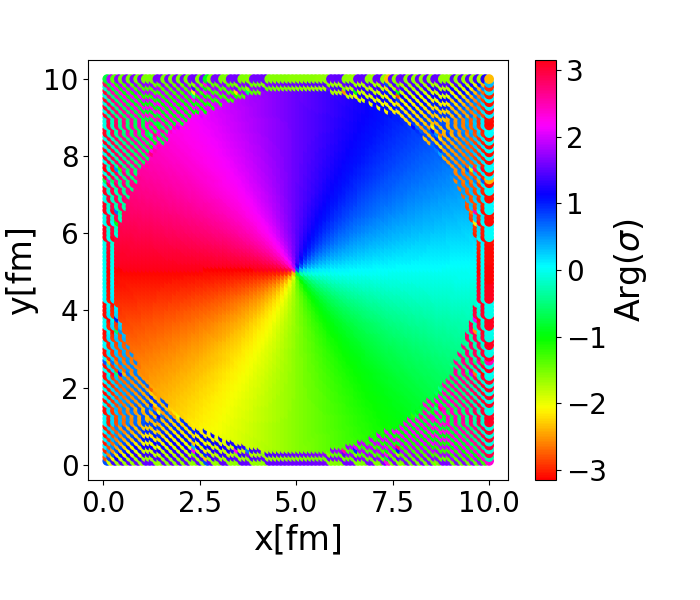}
        \caption{$\arg[\sigma(\mathbf{r})]$}
        \label{fig:nopin_0d0_15d-2_GS_phase_s}
    \end{subfigure}
    \vspace{0.5em}
    \begin{subfigure}[tb]{0.8\columnwidth}
        \centering
        \includegraphics[width=\textwidth]{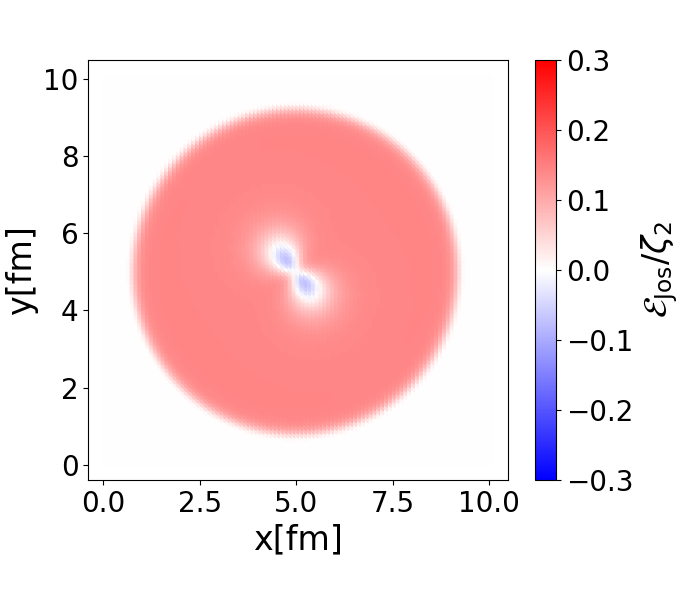}
        \caption{$\mathcal{E}_{\mathrm{Jos}}(\mathbf{r})/\zeta_2$}
        \label{fig:nopin_0d0_15d-2_GS_Rabi}
    \end{subfigure}
    \caption{
        Same as Fig.~\ref{fig:nopin_all_1d-2}, but for $\zeta_1=0$ and
        $\tilde\zeta_2=0.15$.
    }\label{fig:nopin_all_15d-2}
\end{figure}

\begin{figure}[tbp]
    \begin{subfigure}[b]{0.48\columnwidth}
        \centering
        \includegraphics[width=\textwidth]{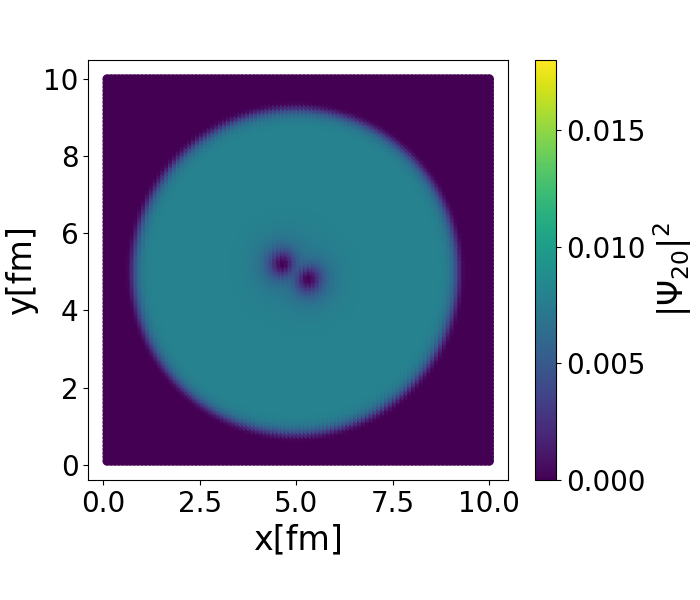}
        \caption{$|\Psi_{20}(\mathbf{r})|^2$}
        \label{fig:nopin_0d0_100d-2_GS_Psi20}
    \end{subfigure}
    \hfill
    \begin{subfigure}[b]{0.48\columnwidth}
        \centering
        \includegraphics[width=\textwidth]{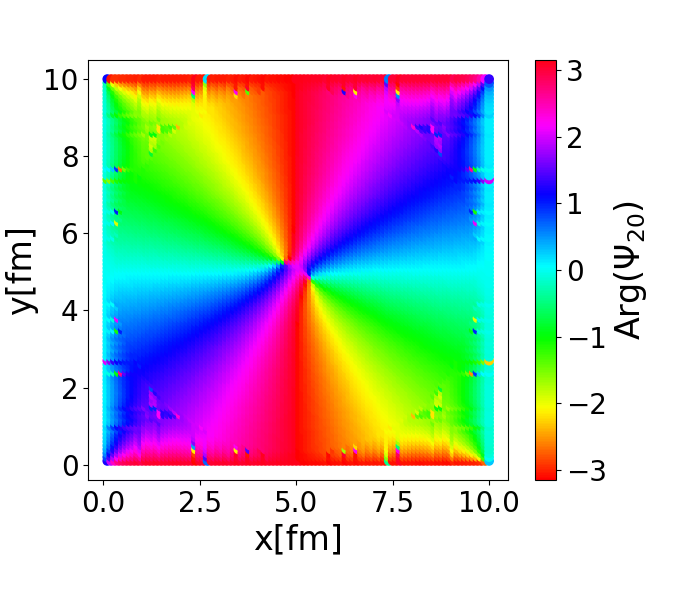}
        \caption{$\arg[\Psi_{20}(\mathbf{r})]$}
        \label{fig:nopin_0d0_100d-2_GS_phase_Psi20}
    \end{subfigure}
    \begin{subfigure}[b]{0.48\columnwidth}
        \centering
        \includegraphics[width=\textwidth]{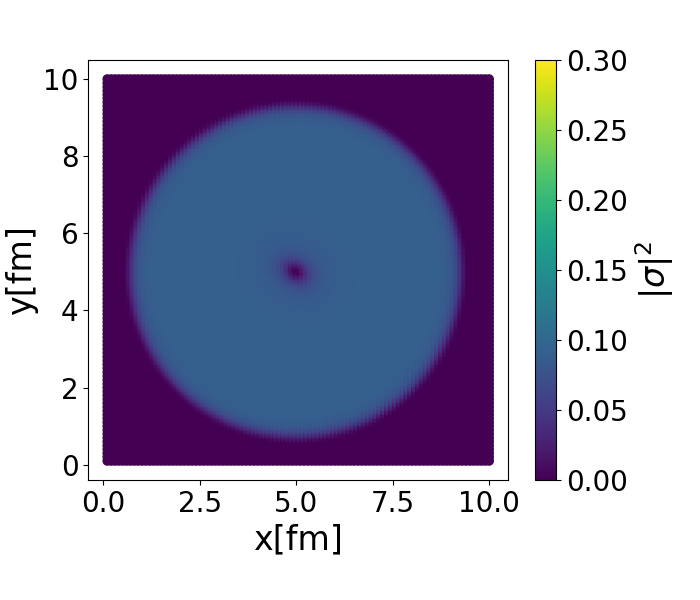}
        \caption{$|\sigma(\mathbf{r})|^2$}
        \label{fig:nopin_0d0_100d-2_GS_rho_s}
    \end{subfigure}
    \hfill
    \begin{subfigure}[b]{0.48\columnwidth}
        \centering
        \includegraphics[width=\textwidth]{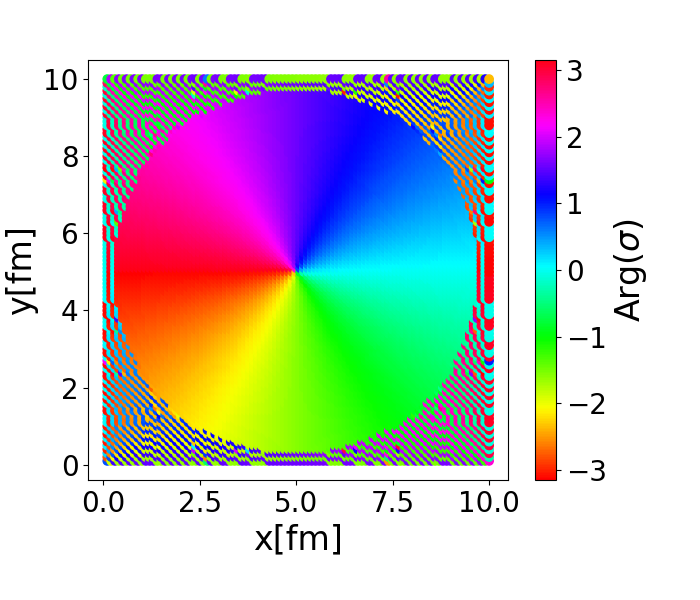}
        \caption{$\arg[\sigma(\mathbf{r})]$}
        \label{fig:nopin_0d0_100d-2_GS_phase_s}
    \end{subfigure}
    \vspace{0.5em}
    \begin{subfigure}[tb]{0.8\columnwidth}
        \centering
        \includegraphics[width=\textwidth]{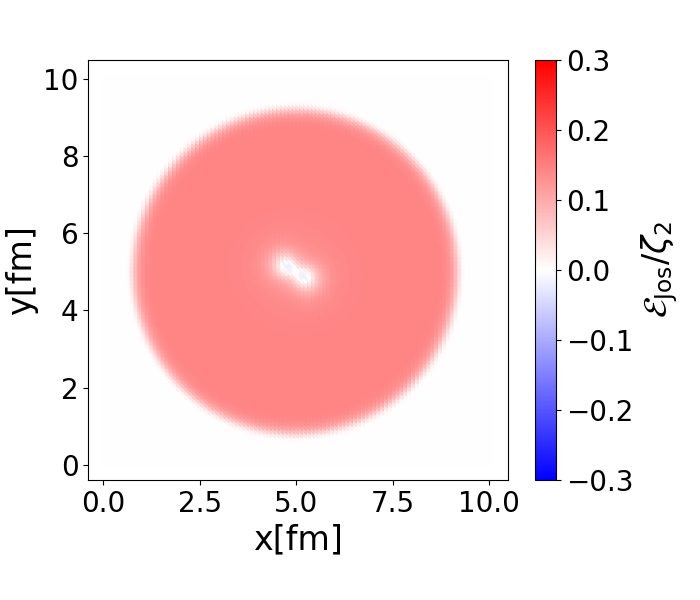}
        \caption{$\mathcal{E}_{\mathrm{Jos}}(\mathbf{r})/\zeta_2$}
        \label{fig:nopin_0d0_100d-2_GS_Rabi}
    \end{subfigure}
    \caption{
        Same as Figs.~\ref{fig:nopin_all_1d-2} and \ref{fig:nopin_all_15d-2}, but for $\zeta_1=0$ and $\tilde\zeta_2=1$.
    }\label{fig:nopin_all_100d-2}
\end{figure}

To get deeper insight into the observed behavior, we show in
Figs.~\ref{fig:nopin_all_1d-2}, \ref{fig:nopin_all_15d-2}, 
and~\ref{fig:nopin_all_100d-2} the actual spatial configurations
of the HQVs and SQV, along with the corresponding distribution of
$\mathcal{E}_{\mathrm{Jos}}(\mathbf{r})$, at the values of
$\tilde\zeta_2$ indicated by the vertical black dotted lines in 
Figs.~\ref{fig:no_trap}(a) and~\ref{fig:no_trap}(b).

Figure~\ref{fig:nopin_all_1d-2} shows the structure of the quantized
superfluid  vortices for $\tilde\zeta_1=0$ and $\tilde\zeta_2=0.01$. 
Figure~\ref{fig:nopin_0d0_1d-2_GS_Psi20} shows the density profile 
$|\Psi_{20}(\mathbf{r})|^2$ of the $^3\text{P}_2$ order parameter,
and  Fig.~\ref{fig:nopin_0d0_1d-2_GS_phase_Psi20} shows its phase 
$\mathrm{arg}[\Psi_{20}(\mathbf{r})]$. Figure~\ref{fig:nopin_0d0_1d-2_GS_rho_s}
shows the density profile $|\sigma(\mathbf{r})|^2$ of the $^1\text{S}_0$
order parameter, and Fig.~\ref{fig:nopin_0d0_1d-2_GS_phase_s} shows
its phase $\mathrm{arg}[\sigma(\mathbf{r})]$. Figure~\ref{fig:nopin_0d0_1d-2_GS_Rabi}
shows the Josephson energy density divided by $\zeta_2$, \textit{i.e.},
$\mathcal{E}_{\mathrm{Jos}}(\mathbf{r})/\zeta_2$, which reflects
the relative phase between the $^1\text{S}_0$ and $^3\text{P}_2$
condensates. A single SQV of the $^1\text{S}_0$ condensate is located
at the center, identified as the point where the density vanishes
(the dark region at the center) in Fig.~\ref{fig:nopin_0d0_1d-2_GS_rho_s}
and around which the phase winds by $2\pi$ as shown in
Fig.~\ref{fig:nopin_0d0_1d-2_GS_phase_s}. The two $^3\text{P}_2$
HQVs, identified analogously from the density-depleted points
in Fig.~\ref{fig:nopin_0d0_1d-2_GS_Psi20} and the surrounding
$2\pi$ phase winding in Fig.~\ref{fig:nopin_0d0_1d-2_GS_phase_Psi20}, 
are located point symmetrically with respect to the SQV.

In Fig.~\ref{fig:nopin_0d0_1d-2_GS_Rabi}, we show the normalized Josephson
energy density, $\mathcal{E}_\text{Jos}(\mathbf{r})/\zeta_2$, to investigate
the effects of the Josephson coupling. In the figure, the red colored region
corresponds to $\cos[2\Delta\theta_m(\mathbf{r})]>0$ and $\Delta\theta_m\approx0$,
while the blue colored region corresponds to $\cos[2\Delta\theta_m(\mathbf{r})]<0$
and $\Delta\theta_m\approx\pi/2$. Since the phase of $\sigma(\mathbf{r})$ rotates by
$2\pi$ around the center and that of $\Psi_{20}(\mathbf{r})$ rotates around each of
the two HQVs, there appear some domains with non-zero phase difference
between $^1\text{S}_0$ and $^3\text{P}_2$ order parameters. Indeed, a
pronounced variation and sign changes of the Josephson energy are observed
in the vicinity of the vortex cores, where the phase fields $\theta_\sigma
(\mathbf{r})$ and $\theta_m(\mathbf{r})$ wind rapidly. As a result,
we find blue colored regions in Fig.~\ref{fig:nopin_0d0_1d-2_GS_Rabi}
between the $^1\text{S}_0$ SQV and the two $^3\text{P}_2$ HQVs.
This blue colored regions between a $^1\text{S}_0$ SQV and a
$^3\text{P}_2$ HQV can be regarded as a domain wall, across which
the relative phase $\Delta\theta_m(\mathbf{r})$ changes abruptly from $0$ to $\pi/2$.
In the present case, there are two domain walls between the SQV and the
two HQVs. This spatial structure reflects the relative phase locking
($\Delta\theta_m\approx 0$) in the outer region and unlocking ($\Delta
\theta_m\neq 0$) in between SQV and HQVs. The sign change of $\mathcal{E}
_{\mathrm{Jos}}(\mathbf{r})$ across the vortex core indicates a local reversal of
the effective Josephson coupling. As can be inferred from Eq.~\eqref{eq:Eden_Jos},
having the blue colored domains is energetically unfavorable, which we
identify as a key driver of the inter-vortex force between the $^1\text{S}_0$
and $^3\text{P}_2$ quantized vortices.

To show such interaction dependence of vortex configurations, we show
vortex configurations obtained for different Josephson coupling strength,
$\tilde\zeta_2=0.15$ and $\tilde\zeta_2=1.00$, in Figs.~\ref{fig:nopin_0d0_15d-2_GS_Psi20}--\ref{fig:nopin_0d0_15d-2_GS_Rabi} and 
Figs.~\ref{fig:nopin_0d0_100d-2_GS_Psi20}--\ref{fig:nopin_0d0_100d-2_GS_Rabi}, 
respectively. The meaning of each panel is the same as Fig.~\ref{fig:nopin_all_1d-2}.
The results shown in 
Figs.~\ref{fig:nopin_0d0_15d-2_GS_Psi20}--\ref{fig:nopin_0d0_15d-2_GS_phase_s} and
Figs.~\ref{fig:nopin_0d0_100d-2_GS_Psi20}--\ref{fig:nopin_0d0_100d-2_GS_phase_s}
very similar to those in 
Figs.~\ref{fig:nopin_0d0_1d-2_GS_Psi20}--\ref{fig:nopin_0d0_1d-2_GS_phase_s},
except slight decrease of vortex distance between two HQVs.
In the case of $\tilde\zeta_2=0.01$ (Fig.~\ref{fig:nopin_0d0_1d-2_GS_Rabi}),
the sign-reversed regions of the Josephson energy density (the blue colored regions)
have a finite width and length. As $\tilde\zeta_2$ increases to $0.15$
(Fig.~\ref{fig:nopin_0d0_15d-2_GS_Rabi}), this region becomes narrower 
and shorter, and for $\tilde\zeta_2=1.00$ (Fig.~\ref{fig:nopin_0d0_100d-2_GS_Rabi}),
it has essentially disappeared.  Correspondingly, the distance between the
two HQVs also decreases with increasing $\tilde\zeta_2$: at $\tilde\zeta_2=0.15$ 
(Fig.~\ref{fig:nopin_0d0_15d-2_GS_Psi20}) the separation is roughly half 
of that at $\tilde\zeta_2=0.01$ 
(Fig.~\ref{fig:nopin_0d0_1d-2_GS_Psi20}), and at $\tilde\zeta_2=1.00$ 
(Fig.~\ref{fig:nopin_0d0_100d-2_GS_Psi20}) the two HQVs are nearly adjacent
to each other. This result demonstrates that the Josephson coupling $\zeta_2$
acts as  a tunable parameter controlling the effective Josephson-like
force between the $^1\text{S}_0$ SQV and the $^3\text{P}_2$ HQVs, with
direct implications for the vortex depinning dynamics relevant to the
glitch mechanism. This alignment indicates that a configuration in which
the two HQVs and one SQV are co-located is energetically more favorable
than one in which the HQVs are spatially separated from the SQV, since
the former minimizes the total Josephson energy. As a result, the two
HQVs and the SQV interact attractively, and the system evolves toward
a bound vortex configuration with increasing $\zeta_2$.

\subsection{Simulation with Pinning Site}\label{SubSec:results_pin}

\begin{figure}\centering
\begin{subfigure}{0.48\textwidth}
    \centering
    \includegraphics[width=0.85\linewidth]{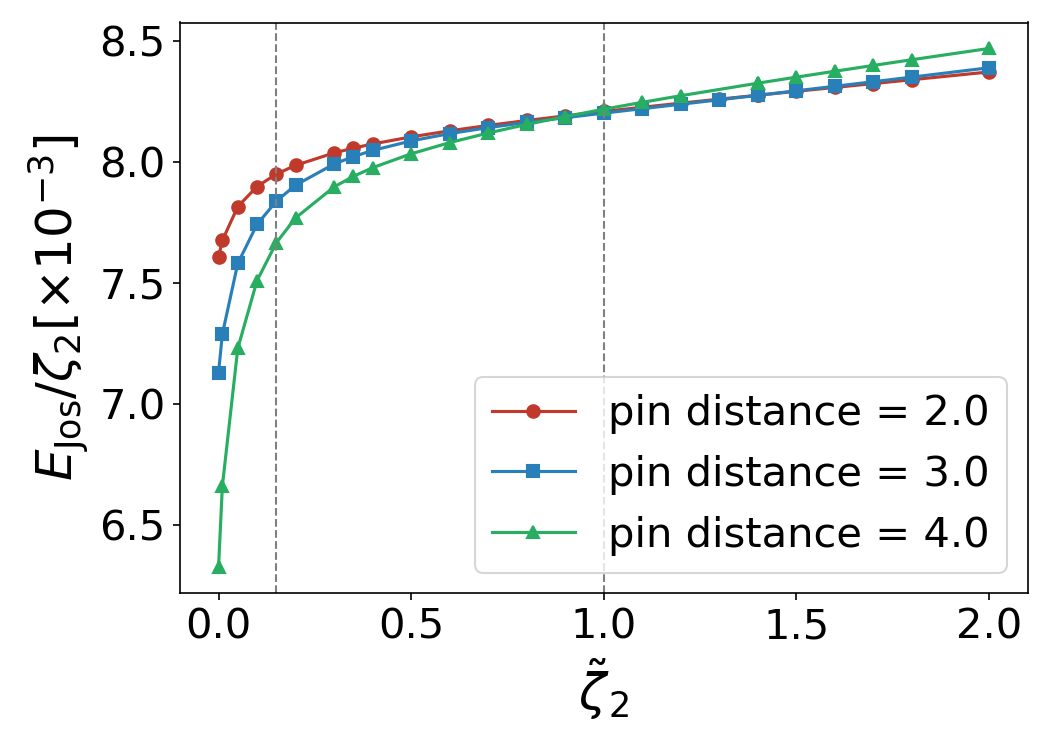}
    \caption{$E_{\mathrm{Jos}}/\zeta_2$}
    \label{fig:Rabi_pinned_per_zeta2}
\end{subfigure}
\begin{subfigure}{0.48\textwidth}
    \centering
    \includegraphics[width=0.85\linewidth]{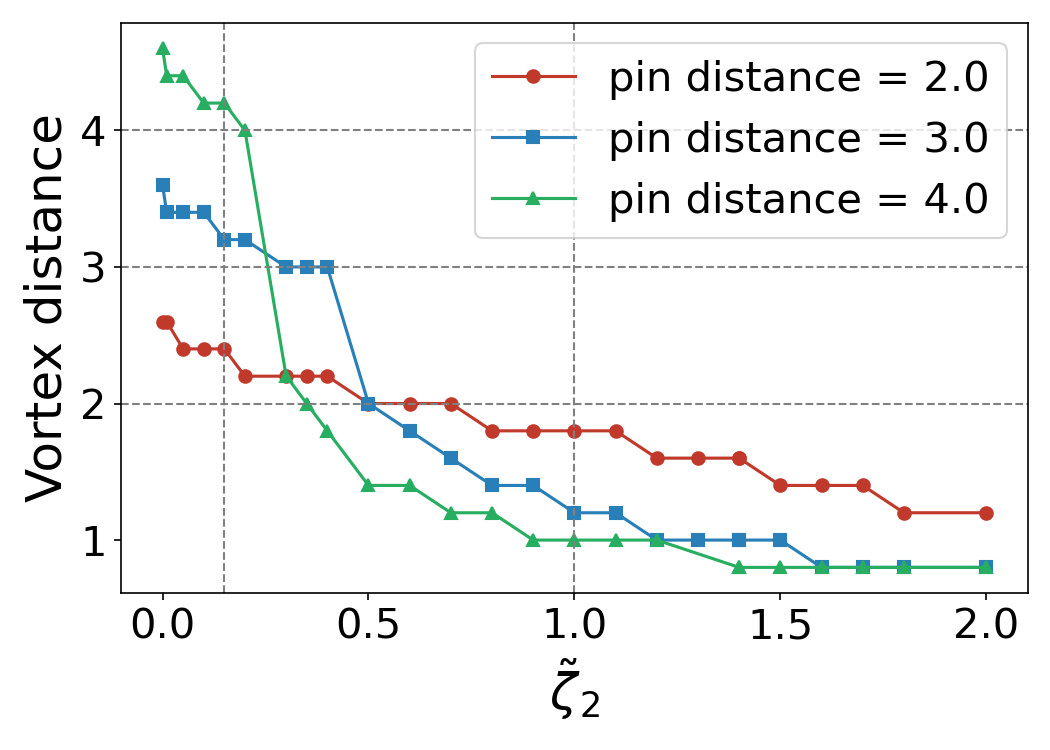}
    \caption{HQVs vortex distance}
    \label{fig:Rabi_pinned_distance}
\end{subfigure}
\caption{
(a)~Josephson energy, $E_\text{Jos}(\mathbf{r})/\zeta_2$, as a function 
    of the Josephson coupling constant $\tilde{\zeta}_2$ for three values of 
    the pinning site distances of HQV, as indicated in the legend. Here, $\zeta_1=0$.
    The system consists of a single SQV in the $^1\text{S}_0$ condensate and two HQVs 
    in the $^3\text{P}_2$ condensate, with singlet and triplet pinning position.
(b)~Inter-vortex distance between the two HQVs in the $^3\text{P}_2$ 
    condensate as a function of the Josephson coupling constant $\tilde{\zeta}_2$ 
    for three values of the pinning site of HQV, 
    as indicated in the legend.
    The system configuration and imaginary-time evolution procedure 
    are the same as in Fig.~\ref{fig:Rabi_pinned_per_zeta2}.
    In both panels, the results for HQV distance of 2.0, 3.0, and 4.0\,fm are
    shown by red filled circles connected with solid lines, blue filled squares
    conneted with solid lines, and green filled triangles connected with solid
    lines, respectively.
    }\label{fig:pinned}
\end{figure}


In the previous section, it is found that the Josephson coupling induces
the effective attractive interaction between SQV and HQVs that colocate them
at the same position. Such attractive interaction may significantly affect
the vortex pinning mechanism near the crust-core boundary. To investigate
how the Josephson coupling affects pinning configurations, here we introduce
artificial pinning potentials for $^1\text{S}_0$ SQV and $^3\text{P}_2$ HQVs.
We set a pinning potential, defined in Sec.~\ref{Sec:method_pin}, for SQV at
the center of the container, while two pinning potentials are located at
$(\pm\frac{d_{\text{pin}}}{2},0)$ with $d_{\text{pin}}=2$, $3$, and $4$\,fm.
We perform calculations for various values of the Josephson coupling constant,
$0\le\tilde\zeta_2\le2$, keeping $\zeta_1=0$ for simplicity, to analyze when
vortices unpinning from the pinning site, and check the vortices structures
during unpinning.

\subsubsection{Energy and Vortex Distance}\label{subsec:energy_distance_pin}

Figure~\ref{fig:Rabi_pinned_per_zeta2} shows the $\zeta_2$-dependence of 
the Josephson term energy normalized by $\zeta_2$, $E_{\mathrm{Jos}}/\zeta_2$, 
for three different pinning-site separations between the two HQVs: $2.0$~fm
(red), $3.0$~fm (blue), and $4.0$~fm (green). Figure~\ref{fig:Rabi_pinned_distance}
shows the corresponding change in the actual inter-vortex distance of the
two HQVs as $\zeta_2$ is varied, using the same color scheme as in
Fig.~\ref{fig:Rabi_pinned_per_zeta2}. 

As seen in Fig.~\ref{fig:Rabi_pinned_per_zeta2}, the Josephson energy
term behaves similarly regardless of the pinning-site separation: 
it increases sharply with $\zeta_2$, and then grows nearly linearly 
beyond $\tilde\zeta_2\simeq0.25$. We note that the magnitude of
$E_\text{Jos}/\zeta$ can not directly be compared with that shown
in Fig.~\ref{fig:Rabi_no_trap_distance}, because the density distribution
is substantially modified due to the presence of three pinning potentials.
Namely, the pinning potentials expel the densities and the condensate densities
in the outer region are kept nearly constant even when vortices are unpinned
with increasing $\tilde\zeta_2$, which results in the monotonic increase of
$E_\text{Jos}/\zeta_2$ in Fig.~\ref{fig:Rabi_pinned_distance}.

As shown in Fig.~\ref{fig:Rabi_pinned_distance}, when $\zeta_2$ is increased 
beyond a critical value, the HQVs are no longer confined to their pinning
sites and are instead attracted toward the SQV. The critical value depends
on the pinning-site separation which is found to be:
$\tilde\zeta_2=0.20$ and $0.50$ for a pinning-site separation of
$d_\text{pin}=4.0$\,fm and $3.0$~fm, respectively. However, for the
$d_\text{pin}=2.0$\,fm case, the depinning was not as abrupt, which occurs
gradually around $0.70<\tilde\zeta_2<1.25$, since the initial separation
was already short. The vortices pinned at closer separations require
a larger $\zeta_2$ to unpin, indicating that the effective attractive
force is weaker at shorter inter-vortex distances. After unpinning,
the resulting inter-vortex distances for the $3.0$\,fm and $4.0$\,fm
cases converge to similar final values. The resulting inter-vortex
distance closely coincides with that obtained in the absence of pinning
potentials (blue squares in Fig.~\ref{fig:Rabi_no_trap_distance}),
confirming that the Josephson-energy-driven attraction dominates
over the pinning force for sufficiently large $\zeta_2$.

\begin{figure}[tbp]
    \begin{subfigure}[b]{0.48\columnwidth}
        \centering
        \includegraphics[width=\textwidth]{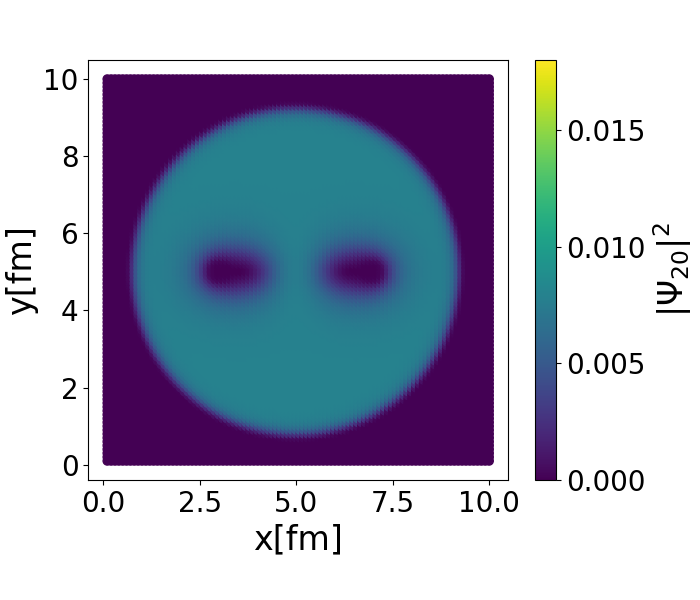}
        \caption{$|\Psi_{20}(\mathbf{r})|^2$}
        \label{fig:GS_Psi20_15d-2}
    \end{subfigure}
    \hfill
    \begin{subfigure}[b]{0.48\columnwidth}
        \centering
        \includegraphics[width=\textwidth]{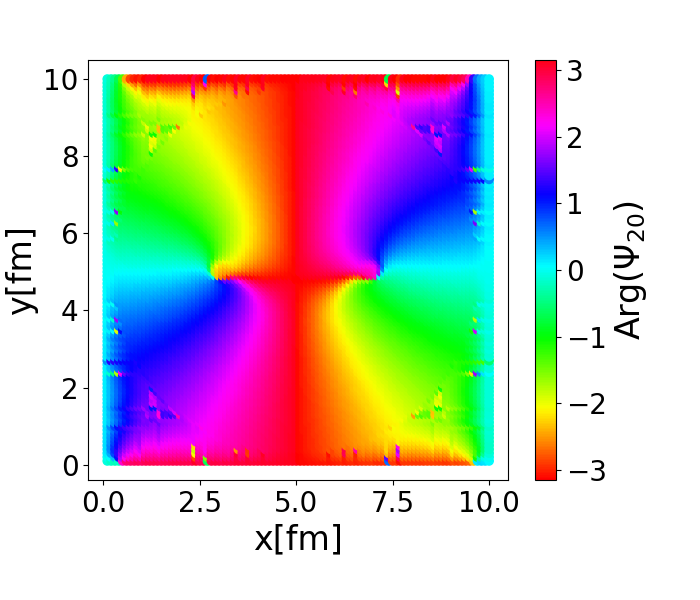}
        \caption{$\arg[\Psi_{20}(\mathbf{r})]$}
        \label{fig:GS_phase_Psi20_15d-2}
    \end{subfigure}
    \begin{subfigure}[b]{0.48\columnwidth}
        \centering
        \includegraphics[width=\textwidth]{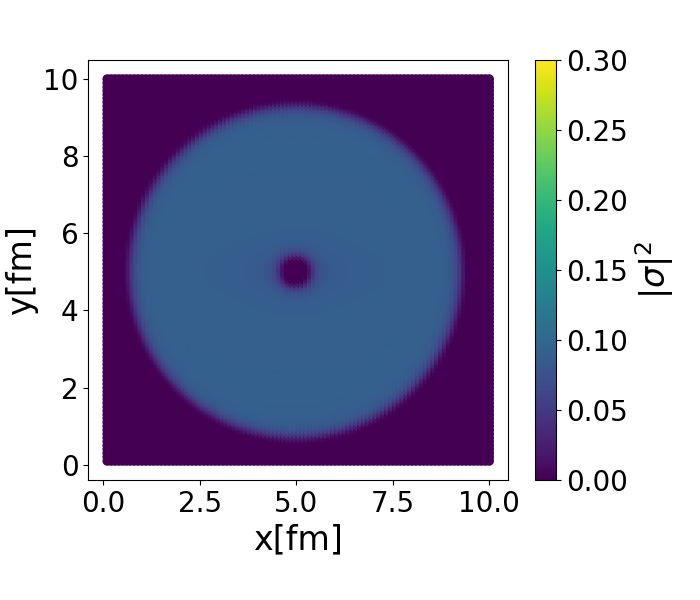}
        \caption{$|\sigma(\mathbf{r})|^2$}
        \label{fig:GS_rho_s_15d-2}
    \end{subfigure}
    \hfill
    \begin{subfigure}[b]{0.48\columnwidth}
        \centering
        \includegraphics[width=\textwidth]{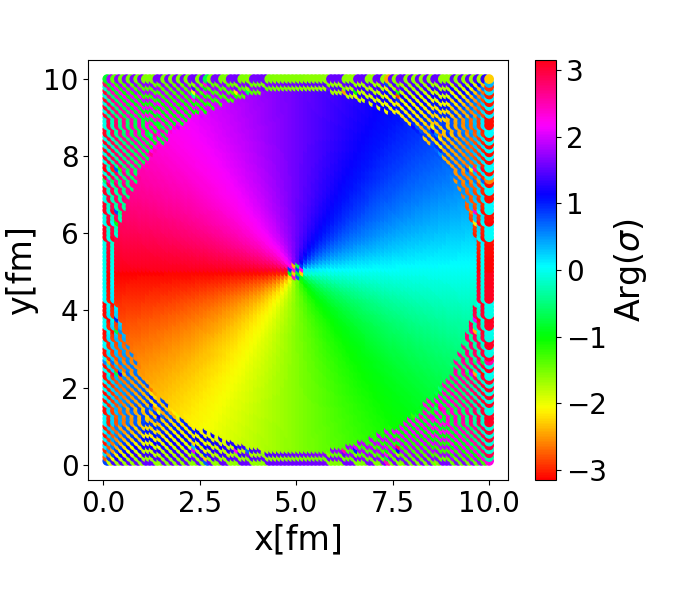}
        \caption{$\arg[\sigma(\mathbf{r})]$}
        \label{fig:GS_phase_s_15d-2}
    \end{subfigure}
    \vspace{0.5em}
    \begin{subfigure}[tb]{0.8\columnwidth}
        \centering
        \includegraphics[width=\textwidth]{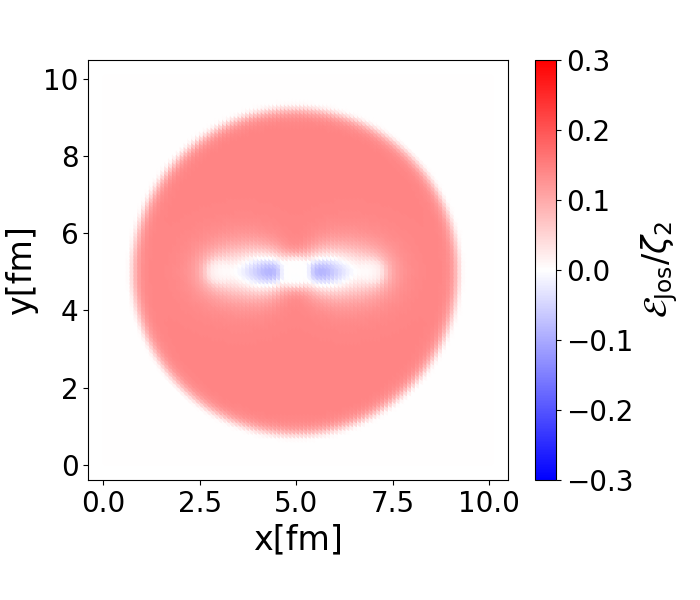}
        \caption{$\mathcal{E}_{\mathrm{Jos}}(\mathbf{r})/\zeta_2$}
        \label{fig:GS_Rabi_withpin_15d-2}
    \end{subfigure}
    \caption{
        Same as Figs.~\ref{fig:nopin_all_1d-2}, \ref{fig:nopin_all_15d-2}, and \ref{fig:nopin_all_100d-2}, but for $\zeta_1=0$ and $\tilde\zeta_2=0.15$, with pinning potentials, where HQV pinning site distance is $d_{\text{pin}}=4.0$~fm.
        }
    \label{fig:withpin_all_15}
\end{figure}

\begin{figure}[tbp]
    \begin{subfigure}[b]{0.48\columnwidth}
        \centering
        \includegraphics[width=\textwidth]{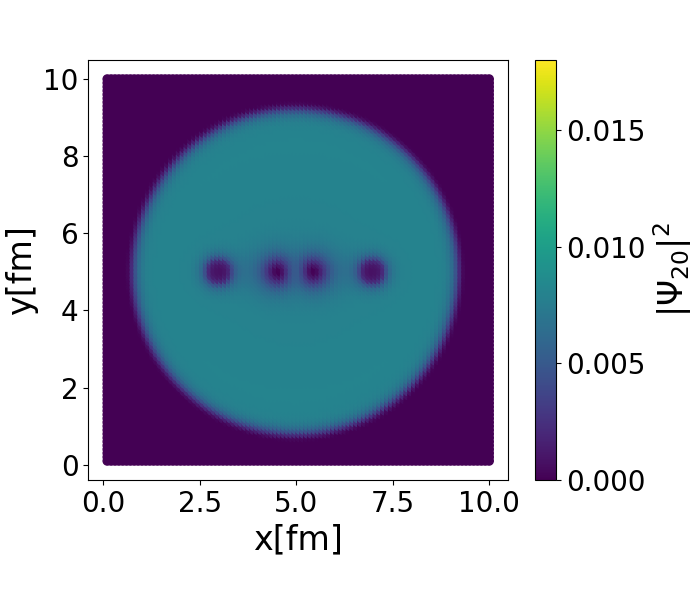}
        \caption{$|\Psi_{20}(\mathbf{r})|^2$}
        \label{fig:GS_Psi20_100d-2}
    \end{subfigure}
    \hfill
    \begin{subfigure}[b]{0.48\columnwidth}
        \centering
        \includegraphics[width=\textwidth]{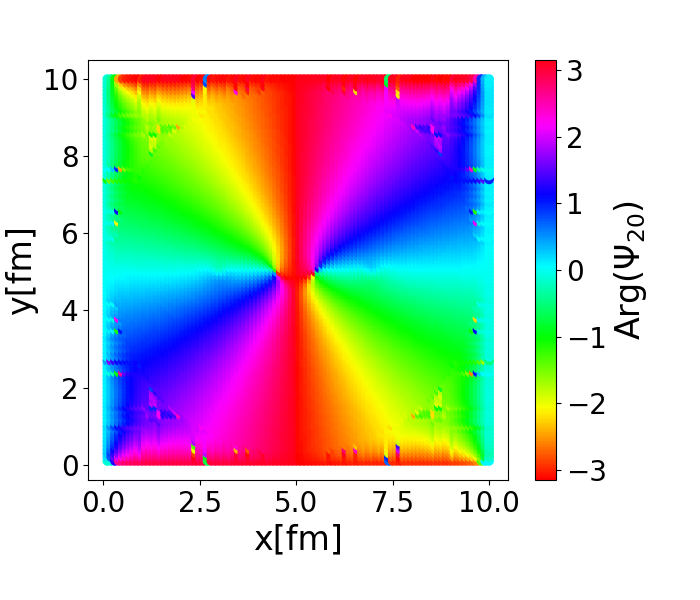}
        \caption{$\arg[\Psi_{20}(\mathbf{r})]$}
        \label{fig:GS_phase_Psi20_100d-2}
    \end{subfigure}
    \begin{subfigure}[b]{0.48\columnwidth}
        \centering
        \includegraphics[width=\textwidth]{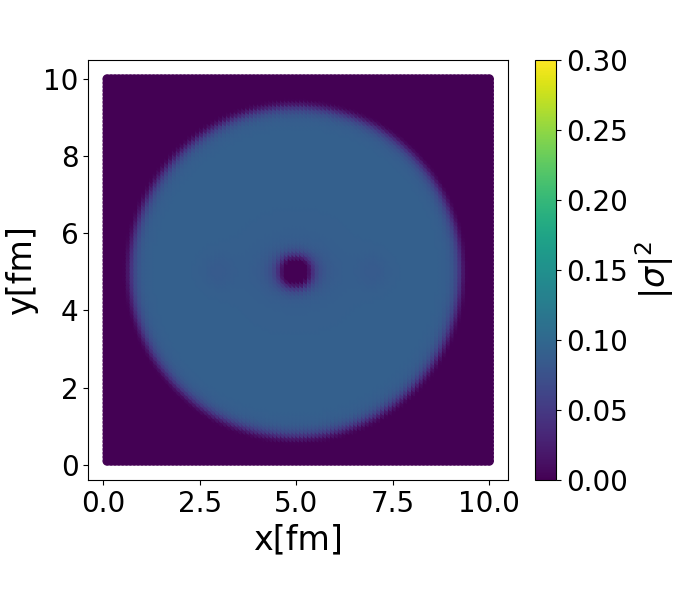}
        \caption{$|\sigma(\mathbf{r})|^2$}
        \label{fig:GS_rho_s_100d-2}
    \end{subfigure}
    \hfill
    \begin{subfigure}[b]{0.48\columnwidth}
        \centering
        \includegraphics[width=\textwidth]{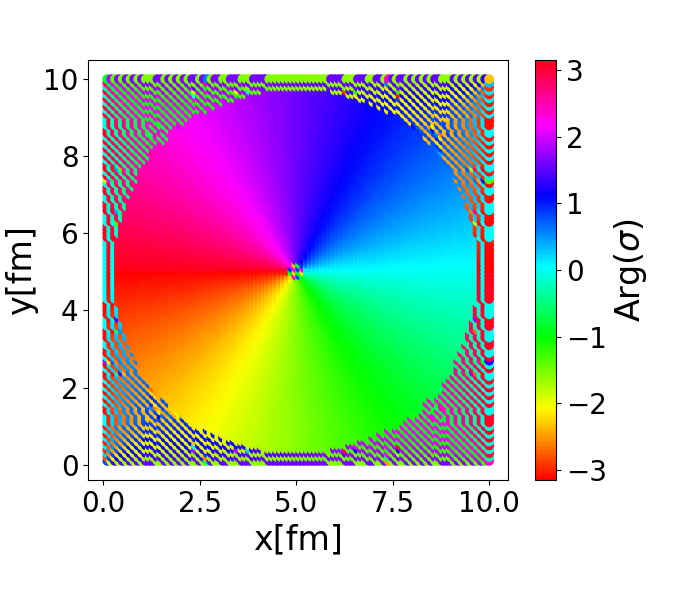}
        \caption{$\arg[\sigma(\mathbf{r})]$}
        \label{fig:GS_phase_s_100d-2}
    \end{subfigure}
    \vspace{0.5em}
    \begin{subfigure}[tb]{0.8\columnwidth}
        \centering
        \includegraphics[width=\textwidth]{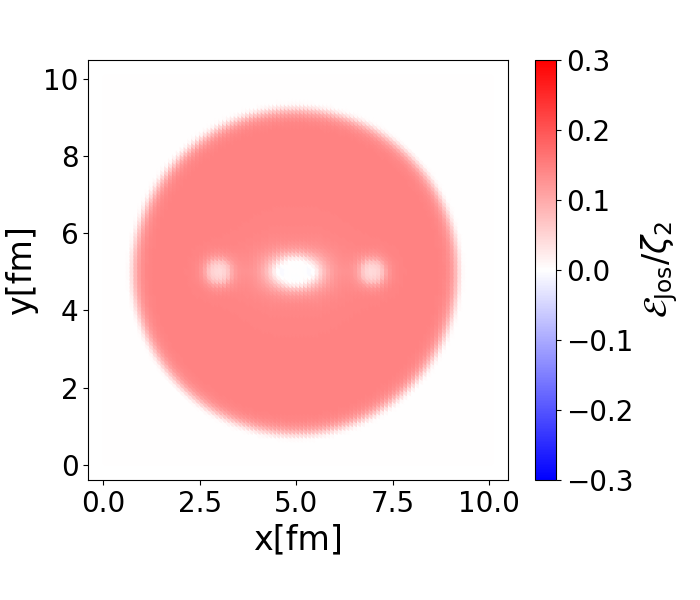}
        \caption{$\mathcal{E}_{\mathrm{Jos}}(\mathbf{r})/\zeta_2$}
        \label{fig:GS_Rabi_withpin_100d-2}
    \end{subfigure}
    \caption{
        Same as Fig.~\ref{fig:withpin_all_15}, but for $\zeta_1=0$ and
        $\tilde\zeta_2=1$.
    }\label{fig:withpin_all_100}
\end{figure}

\subsubsection{Vortex Structure}\label{subsec:structure_pin}

The vortex configurations before and after unpinning, for a pinning-site 
separation of $d_{\text{pin}}=4.0$~fm, are shown in Fig.~\ref{fig:withpin_all_15} 
(before unpinning) and Fig.~\ref{fig:withpin_all_100} (after unpinning).
Each panel of Figs.~\ref{fig:withpin_all_15} and \ref{fig:withpin_all_100}
shows the same quantity as shown in Figs.~\ref{fig:nopin_all_1d-2},
\ref{fig:nopin_all_15d-2}, and~\ref{fig:nopin_all_100d-2}.

From Fig.~\ref{fig:GS_Rabi_withpin_15d-2}, we find that, in the weak-coupling
regime, where $\zeta_2$ is small and the vortices remain pinned, the
domain walls---the region where the sign of the Josephson energy density
is reversed (the blue region in Fig.~\ref{fig:GS_Rabi_withpin_15d-2})---corresponding
to where the phases of the $^3\text{P}_2$ and $^1\text{S}_0$ condensates
are out of phase---are distributed narrowly, so as to minimize
its area. As a result, the phase of $\Psi_{20}(\mathbf{r})$ around
the $^3\text{P}_2$ vortices is significantly distorted, as seen in
Fig.~\ref{fig:GS_phase_Psi20_15d-2}, reflecting the strong attractive
force exerted on the HQVs by the Josephson term. In other words,
the velocity field of $^3\text{P}_2$ superfluid is largely distorted
because of the Josephson coupling.

Once unpinning eventually occurs, the Josephson energy density becomes 
positive throughout the entire domain, as shown in 
Fig.~\ref{fig:GS_Rabi_withpin_100d-2}, indicating that the phases
of the $^3\text{P}_2$ and $^1\text{S}_0$ condensates become aligned. 
Consequently, the two HQVs bind to the SQV after unpinning, forming
a composite vortex structure (\textit{cf.}\ Figs.~\ref{fig:GS_Psi20_100d-2}
and~\ref{fig:GS_rho_s_100d-2}). The resulting inter-vortex distance
in this case is nearly identical to that obtained in the absence of pinning sites.

\section{Summary and prospect}\label{Sec:Conclusion}

In this work, we have investigated the microscopic interaction between 
harf-quantized vortcies (HQVs) in the $^3\text{P}_2$ neutron superfluid and 
a singly-quantized vortex (SQV) in the $^1\text{S}_0$ neutron superfluid 
in a two-dimensional coexistence phase, based on the Gross-Pitaevskii (GP)
framework. We have demonstrated that the Josephson coupling which gives
rise to the Josephson energy term [Eq.~\eqref{eq:Eden_Jos}], induces
a significant attractive interaction between the two $^3\text{P}_2$ HQVs
and one $^1\text{S}_0$ SQV. This attraction is strong enough to overcome
the pinning potential applied to the $^3\text{P}_2$ vortices, driving
them away from their pinning sites and toward the SQV. The resulting
HQV--SQV--HQV bound configuration substantially alters the spatial
arrangement of the $^3\text{P}_2$ vortices, providing microscopic
evidence that the boojum structure proposed in Ref.~\cite{Marmorini2024}
can indeed be formed at the phase boundary between the inner crust and
the outer core of a neutron star. Furthermore, the strong attraction
exerted on the $^1\text{S}_0$ SQV by the $^3\text{P}_2$ HQVs suggests
that, if the triplet vortices are pinned by some mechanism, the singlet
vortex will also experience an attractive force from the triplet vortices,
potentially suppressing the relative motion of the two vortex species 
and affecting the dynamics relevant to the glitch mechanism.

We note that the present study constitutes a first step toward evaluating
the role of the boojum structure in neutron star glitches, \textit{e.g.}
in the vortex network model of Ref.~\cite{Marmorini2024}, microscopically.
The simulations are performed in two spatial dimensions under the assumption
of a coexistence phase, which requires the phase boundary to be planar.
A fully three-dimensional treatment will be needed to capture the realistic
geometry of the inner-crust--outer-core boundary.

The present results motivate several directions for future work. First,
three-dimensional simulations will allow us to evaluate how the HQV bends
in the vicinity of a boojum, and whether the HQV--SQV connection can form
a large-scale vortex network of the kind envisioned in Ref.~\cite{Marmorini2024}.
Three-dimensional calculations are already underway \cite{hattori_sekizawa_2025}.
It is also important to investigate the effects of the Josephson coupling
on vortex lattices, as studied in two-component BEC systems, since the
collective behavior of coupled vortex lattices---in particular, whether
the $^1\text{S}_0$ and $^3\text{P}_2$ vortex lattices lock together or
slide independently---is directly relevant to the angular momentum transfer
mechanism underlying the glitch phenomenon \cite{Cipriani:2013nya}. Second,
the effect of the HQV--SQV attraction on the dynamics of the SQV itself will
be assessed, as the suppression of SQV motion by pinned triplet vortices
may have direct implications for the avalanche model of pulsar glitches.
We note here that we can naturally extend our formalism for real-time
evolution to investigate complex vortices dynamics, including the
mixed phase, under various conditions.

These microscopic results on the vortex--vortex coupling will then be used 
to evaluate the influence of the HQV--SQV attraction on the vortex network
model and the avalanche mechanism, with the aim of providing a quantitative 
microscopic basis for the glitch mechanism. Additionally, the formation of
a stable HQV--SQV--HQV molecular bound state opens the possibility of a novel
oscillation mode arising from the attractive inter-vortex interaction.
Such an inter-vortex vibrational mode could serve as a new internal heat
source in neutron stars, with potential implications for neutron star cooling.
The possibility of recombination and exchange of the HQV--SQV--HQV pair upon
vortex collisions will also be examined in future work. For instance,
exchanging a partner in a collision of two vortex molecules found in
a two-component Bose-Einstein condensation (BEC) \cite{Eto:2019uhe} may occur.

The significant role of the Josephson term found in this work suggests
broader implications beyond neutron star physics. In finite nuclei, both
spin-singlet and spin-triplet pairings are in principle possible, and recent
studies have shown that spin-triplet pairing condensates can coexist with
spin-singlet pairing in even-even singly closed nuclei such as Ca and Sn
isotopes~\cite{Hinohara2024}. The present results indicate that the Josephson
coupling between singlet and triplet condensates---driven by their relative
phase---can generate a strong inter-condensate interaction, suggesting
that analogous Josephson effects may play an important role in finite
nuclei where both $^3\text{P}_2$ and $^1\text{S}_0$ pairings coexist,
as well as in proton--neutron systems coupled via the $^3\text{S}_1$
channel. To quantify these effects in realistic nuclear systems,
it will be necessary to determine the triplet pairing coupling constant 
and the singlet--triplet interaction coefficient from experimental data
on heavy nuclei such as Pb isotopes.

Furthermore, recent microscopic calculations based on superfluid band theory
have indicated the emergence of spin-triplet pairing in the rod (pasta)
phase of the neutron star inner crust, where two-dimensional crystalline
structures, spin-orbit interactions, and strong magnetic fields act in
concert~\cite{Yoshimura2026}. Since the pasta phase is expected to occupy
the high-density region of the inner crust near the crust--core boundary,
the vortex dynamics in the vicinity of the pasta phase may be influenced
by the inter-condensate Josephson coupling identified in the present work,
with potential implications for the glitch mechanism in this transitional region.

Finally, the inter-condensate interaction studied here is directly analogous
to the Josephson coupling in two-component BEC systems, as recently realized
experimentally in bilayer superfluid systems~\cite{Rydow2025,Dominici2015}.
The present framework is therefore also relevant to the broader context of
multi-component condensate physics, including the dynamics of coupled
half-integer and integer vortices in spinor BEC 
systems~\cite{Masuda:2015jka,Kawaguchi2012}.

\section*{Acknowledgments}
This work is supported by JSPS Grant-in-Aid for Scientific Research
KAKENHI Grants No.~JP23K03410 (K.\,S.) and No.~JP23K25864 (K.\,S.),
No.~JP25H01269 (K.\,S.) and JP23K22492 (M.\,N.). 
This work is also supported by JST SPRING, Grant Number JPMJSP 
JPMJSP2180 (T.\,H.) and by the Science Tokyo Support Program for Doctoral Students (T.\,H.), 
funded by the Universities for International Research Excellence.
The work is also supported in part by the WPI program ``Sustainability with Knotted Chiral Meta Matter (WPI-SKCM$^2$)'' at Hiroshima University (M.\,N.). 
This work used computational resources of the Yukawa-21 and Heian supercomputer at Yukawa Institute for Theoretical Physics (YITP), Kyoto University.

\bibliography{main}

@article{Cheng1988,
  author  = {Cheng, K. S. and Pines, D. and Alpar, M. A. and Shaham, J.},
  title   = {Spontaneous Superfluid Unpinning and the Inhomogeneous Distribution of Vortex Lines in Neutron Stars},
  journal = {ApJ},
  volume  = {330},
  pages   = {835},
  year    = {1988},
  doi     = {10.1086/166517}
}

@article{Warszawski2011,
  author  = {Warszawski, L. and Melatos, A.},
  title   = {Gross-Pitaevskii model of pulsar glitches},
  journal = {MNRAS},
  volume  = {415},
  pages   = {1611},
  year    = {2011},
  doi     = {https://doi.org/10.1111/j.1365-2966.2011.18803.x}
}

@article{Ray_2019,
doi = {10.3847/1538-4357/ab24d8},
url = {https://doi.org/10.3847/1538-4357/ab24d8},
year = {2019},
month = {jul},
publisher = {The American Astronomical Society},
volume = {879},
number = {2},
pages = {130},
author = {Ray, Paul S. and Guillot, Sebastien and Ho, Wynn C. G. and Kerr, Matthew and Enoto, Teruaki and Gendreau, Keith C. and Arzoumanian, Zaven and Altamirano, Diego and Bogdanov, Slavko and Campion, Robert and Chakrabarty, Deepto and Deneva, Julia S. and Jaisawal, Gaurava K. and Kozon, Robert and Malacaria, Christian and Strohmayer, Tod E. and Wolff, Michael T.},
title = {Anti-glitches in the Ultraluminous Accreting Pulsar NGC 300 ULX-1 Observed with NICER},
journal = {The Astrophysical Journal}
}

@article{Fujiwara_2024,
doi = {10.1088/1475-7516/2024/03/051},
url = {https://doi.org/10.1088/1475-7516/2024/03/051},
year = {2024},
month = {mar},
publisher = {IOP Publishing},
volume = {2024},
number = {03},
pages = {051},
author = {Fujiwara, Motoko and Hamaguchi, Koichi and Nagata, Natsumi and Ramirez-Quezada, Maura E.},
title = {Vortex creep heating in neutron stars},
journal = {Journal of Cosmology and Astroparticle Physics}
}

@article{Dib_2008,
doi = {10.1086/524653},
url = {https://doi.org/10.1086/524653},
year = {2008},
month = {feb},
publisher = {},
volume = {673},
number = {2},
pages = {1044},
author = {Dib, Rim and Kaspi, Victoria M. and Gavriil, Fotis P.},
title = {Glitches in Anomalous X-Ray Pulsars},
journal = {The Astrophysical Journal}
}

@article{Marmorini2024,
  author  = {Marmorini, G. and Yasui, S. and Nitta, M.},
  title   = {Vortex network in neutron stars},
  journal = {Sci. Rep.},
  volume  = {14},
  pages   = {7857},
  year    = {2024},
  doi     = {https://doi.org/10.1038/s41598-024-56383-w}
}

@article{hattori_sekizawa_2025,
      title={Exploring Interplays Between $^3\text{P}_2$ Neutron Superfluid Vortices and $^1\text{S}_0$ Proton Fluxtubes in the Outer Core of Neutron Stars}, 
      author={Tatsuhiro Hattori and Kazuyuki Sekizawa},
	DOI= "10.1051/epjconf/202637806021",
	url= "https://doi.org/10.1051/epjconf/202637806021",
	journal = {EPJ Web Conf.},
	year = 2026,
	volume = 378,
	pages = "06021",
      eprint={2512.22577},
      archivePrefix={arXiv},
      primaryClass={nucl-th},
}

@article{Drummond2017,
  author  = {Drummond, L. V. and Melatos, A.},
  title   = {Stability of interstitial vortex pinning in a neutron star},
  journal = {MNRAS},
  volume  = {472},
  pages   = {4851},
  year    = {2017},
  doi     = {https://doi.org/10.1093/mnras/stx2301}
}

@article{Leinson2020,
  author  = {Leinson, L. B.},
  title   = {Vortex lattice in the bulk of a neutron star},
  journal = {MNRAS},
  volume  = {498},
  pages   = {304},
  year    = {2020},
  doi     = {https://doi.org/10.1093/mnras/staa2475}
}

@article{Masuda:2015jka,
    author = "Masuda, Kota and Nitta, Muneto",
    title = "{Magnetic Properties of Quantized Vortices in Neutron $^3P_2$ Superfluids in Neutron Stars}",
    eprint = "1512.01946",
    archivePrefix = "arXiv",
    primaryClass = "nucl-th",
    doi = "10.1103/PhysRevC.93.035804",
    journal = "Phys. Rev. C",
    volume = "93",
    number = "3",
    pages = "035804",
    year = "2016"
}

@article{Kobayashi:2022moc,
    author = "Kobayashi, Michikazu and Nitta, Muneto",
    title = "{Core structures of vortices in Ginzburg-Landau theory for neutron $^3P_2$ superfluids}",
    eprint = "2203.09300",
    archivePrefix = "arXiv",
    primaryClass = "nucl-th",
    doi = "10.1103/PhysRevC.105.035807",
    journal = "Phys. Rev. C",
    volume = "105",
    number = "3",
    pages = "035807",
    year = "2022"
}

@article{Kobayashi:2022dae,
    author = "Kobayashi, Michikazu and Nitta, Muneto",
    title = "{Proximity effects of vortices in neutron 3P2 superfluids in neutron stars: Vortex core transitions and covalent bonding of vortex molecules}",
    eprint = "2209.07205",
    archivePrefix = "arXiv",
    primaryClass = "nucl-th",
    doi = "10.1103/PhysRevC.107.045801",
    journal = "Phys. Rev. C",
    volume = "107",
    number = "4",
    pages = "045801",
    year = "2023"
}

@article{Masaki:2019rsz,
    author = "Masaki, Yusuke and Mizushima, Takeshi and Nitta, Muneto",
    title = "{Microscopic description of axisymmetric vortices in $^{3}P_{2}$ superfluids}",
    eprint = "1908.06215",
    archivePrefix = "arXiv",
    primaryClass = "cond-mat.supr-con",
    doi = "10.1103/PhysRevResearch.2.013193",
    journal = "Phys. Rev. Res.",
    volume = "2",
    number = "1",
    pages = "013193",
    year = "2020"
}

@article{Masaki:2021hmk,
    author = "Masaki, Yusuke and Mizushima, Takeshi and Nitta, Muneto",
    title = "{Non-Abelian half-quantum vortices in 3P2 topological superfluids}",
    eprint = "2107.02448",
    archivePrefix = "arXiv",
    primaryClass = "cond-mat.supr-con",
    doi = "10.1103/PhysRevB.105.L220503",
    journal = "Phys. Rev. B",
    volume = "105",
    number = "22",
    pages = "L220503",
    year = "2022"
}

@inproceedings{Masaki:2023rtn,
    author = "Masaki, Yusuke and Mizushima, Takeshi and Nitta, Muneto",
    title = "{Non-Abelian Anyons and Non-Abelian Vortices in Topological Superconductors}",
    eprint = "2301.11614",
    archivePrefix = "arXiv",
    primaryClass = "cond-mat.supr-con",
    doi = "10.1016/B978-0-323-90800-9.00225-0",
booktitle = "Encyclopedia of Condensed Matter Physics (Second Edition)",
Volume = "2", 
pages = "755-794", 
publisher = "Elsevier",
year = "2024"
}

@article{Mizushima:2016fbn,
    author = "Mizushima, Takeshi and Masuda, Kota and Nitta, Muneto",
    title = "{$^3P_2$ superfluids are topological}",
    eprint = "1607.07266",
    archivePrefix = "arXiv",
    primaryClass = "cond-mat.supr-con",
    doi = "10.1103/PhysRevB.95.140503",
    journal = "Phys. Rev. B",
    volume = "95",
    number = "14",
    pages = "140503",
    year = "2017"
}

@article{Mizushima:2019spl,
    author = "Mizushima, Takeshi and Yasui, Shigehiro and Nitta, Muneto",
    title = "{Critical end point and universality class of neutron $^3P_2$ superfluids in neutron stars}",
    eprint = "1908.07944",
    archivePrefix = "arXiv",
    primaryClass = "nucl-th",
    doi = "10.1103/PhysRevResearch.2.013194",
    journal = "Phys. Rev. Res.",
    volume = "2",
    number = "1",
    pages = "013194",
    year = "2020"
}

@article{Mizushima:2021qrz,
    author = "Mizushima, Takeshi and Yasui, Shigehiro and Inotani, Daisuke and Nitta, Muneto",
    title = "{Spin-polarized phases of P23 superfluids in neutron stars}",
    eprint = "2108.01256",
    archivePrefix = "arXiv",
    primaryClass = "nucl-th",
    doi = "10.1103/PhysRevC.104.045803",
    journal = "Phys. Rev. C",
    volume = "104",
    number = "4",
    pages = "045803",
    year = "2021"
}

@article{Yasui:2020xqb,
    author = "Yasui, Shigehiro and Inotani, Daisuke and Nitta, Muneto",
    title = "{Coexistence phase of $^{1}S_{0}$ and $^{3}P_{2}$ superfluids in neutron stars}",
    eprint = "2002.05429",
    archivePrefix = "arXiv",
    primaryClass = "nucl-th",
    doi = "10.1103/PhysRevC.101.055806",
    journal = "Phys. Rev. C",
    volume = "101",
    number = "5",
    pages = "055806",
    year = "2020"
}

@article{Kobayashi2021,
  author = {Kobayashi, Michikazu and Nitta, Muneto},
  title = {Symmetry classification of uniform states in spin-2 Bose-Einstein condensates and neutron $^{3}P_{2}$ superfluids},
  journal = {Phys. Rev. A},
  volume = {104},
  issue = {5},
  pages = {053302},
  numpages = {12},
  year = {2021},
  month = {Nov},
  publisher = {American Physical Society},
  doi = {10.1103/PhysRevA.104.053302},
  url = {https://link.aps.org/doi/10.1103/PhysRevA.104.053302}
}

@article{Kawaguchi2012,
title = {Spinor Bose–Einstein condensates},
journal = {Physics Reports},
volume = {520},
number = {5},
pages = {253-381},
year = {2012},
issn = {0370-1573},
doi = {https://doi.org/10.1016/j.physrep.2012.07.005},
url = {https://www.sciencedirect.com/science/article/pii/S0370157312002098},
author = {Yuki Kawaguchi and Masahito Ueda}
}

@article{Ueda2002,
  author  = {Koashi, M. and Ueda, M.},
  title   = {Exact Eigenstates and Magnetic Response of Spin-1 and Spin-2 
             {Bose-Einstein} Condensates},
  journal = {Phys. Rev. Lett.},
  volume  = {84},
  pages   = {1066},
  year    = {2000},
  doi     = {10.1103/PhysRevLett.84.1066}
}

@article{Rydow2025,
  author  = {Rydow, E. and Singh, V. P. and Beregi, A. and Chang, E. 
             and Mathey, L. and Foot, C. J. and Sunami, S.},
  title   = {Observation of a bilayer superfluid with interlayer coherence},
  journal = {Nat. Commun.},
  volume  = {16},
  pages   = {7201},
  year    = {2025},
  doi     = {10.1038/s41467-025-62277-w}
}

@article{Dominici2015,
  author  = {Dominici, L. and Dagvadorj, G. and Fellows, J. M. 
             and Ballarini, D. and De Giorgi, M. and Marchetti, F. M.
             and Piccirillo, B. and Marrucci, L. and Bramati, A. 
             and Gigli, G. and Szyma{\'n}ska, M. H. and Sanvitto, D.},
  title   = {Vortex and half-vortex dynamics in a nonlinear spinor quantum fluid},
  journal = {Sci. Adv.},
  volume  = {1},
  pages   = {e1500807},
  year    = {2015},
  doi     = {10.1126/sciadv.1500807}
}

@article{Lattimer2016,
title = {The equation of state of hot, dense matter and neutron stars},
journal = {Physics Reports},
volume = {621},
pages = {127-164},
year = {2016},
note = {Memorial Volume in Honor of Gerald E. Brown},
issn = {0370-1573},
doi = {https://doi.org/10.1016/j.physrep.2015.12.005},
url = {https://www.sciencedirect.com/science/article/pii/S0370157315005396},
author = {James M. Lattimer and Madappa Prakash}
}

@Inbook{Haskell2018,
author="Haskell, Brynmor
and Sedrakian, Armen",
editor="Rezzolla, Luciano
and Pizzochero, Pierre
and Jones, David Ian
and Rea, Nanda
and Vida{\~{n}}a, Isaac",
title="Superfluidity and Superconductivity in Neutron Stars",
bookTitle="The Physics and Astrophysics of Neutron Stars",
year="2018",
publisher="Springer International Publishing",
address="Cham",
pages="401--454",
isbn="978-3-319-97616-7",
doi="10.1007/978-3-319-97616-7{\_}8",
}

@article{Ozel2016,
   author = "Özel, Feryal and Freire, Paulo",
   title = "Masses, Radii, and the Equation of State of Neutron Stars", 
   journal= "Annual Review of Astronomy and Astrophysics",
   year = "2016",
   volume = "54",
   number = "Volume 54, 2016",
   pages = "401-440",
   doi = "https://doi.org/10.1146/annurev-astro-081915-023322",
   url = "https://www.annualreviews.org/content/journals/10.1146/annurev-astro-081915-023322",
   publisher = "Annual Reviews",
   issn = "1545-4282",
   type = "Journal Article",
}

@article{Miller2021,
doi = {10.3847/2041-8213/ac089b},
url = {https://doi.org/10.3847/2041-8213/ac089b},
year = {2021},
month = {sep},
publisher = {The American Astronomical Society},
volume = {918},
number = {2},
pages = {L28},
author = {Miller, M. C. and Lamb, F. K. and Dittmann, A. J. and Bogdanov, S. and Arzoumanian, Z. and Gendreau, K. C. and Guillot, S. and Ho, W. C. G. and Lattimer, J. M. and Loewenstein, M. and Morsink, S. M. and Ray, P. S. and Wolff, M. T. and Baker, C. L. and Cazeau, T. and Manthripragada, S. and Markwardt, C. B. and Okajima, T. and Pollard, S. and Cognard, I. and Cromartie, H. T. and Fonseca, E. and Guillemot, L. and Kerr, M. and Parthasarathy, A. and Pennucci, T. T. and Ransom, S. and Stairs, I.},
title = {The Radius of PSR J0740+6620 from NICER and XMM-Newton Data},
journal = {The Astrophysical Journal Letters}
}

@article{Riley2021,
doi = {10.3847/2041-8213/ac0a81},
url = {https://doi.org/10.3847/2041-8213/ac0a81},
year = {2021},
month = {sep},
publisher = {The American Astronomical Society},
volume = {918},
number = {2},
pages = {L27},
author = {Riley, Thomas E. and Watts, Anna L. and Ray, Paul S. and Bogdanov, Slavko and Guillot, Sebastien and Morsink, Sharon M. and Bilous, Anna V. and Arzoumanian, Zaven and Choudhury, Devarshi and Deneva, Julia S. and Gendreau, Keith C. and Harding, Alice K. and Ho, Wynn C. G. and Lattimer, James M. and Loewenstein, Michael and Ludlam, Renee M. and Markwardt, Craig B. and Okajima, Takashi and Prescod-Weinstein, Chanda and Remillard, Ronald A. and Wolff, Michael T. and Fonseca, Emmanuel and Cromartie, H. Thankful and Kerr, Matthew and Pennucci, Timothy T. and Parthasarathy, Aditya and Ransom, Scott and Stairs, Ingrid and Guillemot, Lucas and Cognard, Ismael},
title = {A NICER View of the Massive Pulsar PSR J0740+6620 Informed by Radio Timing and XMM-Newton Spectroscopy},
journal = {The Astrophysical Journal Letters}
}

@article{Page2011,
  title = {Rapid Cooling of the Neutron Star in Cassiopeia A Triggered by Neutron Superfluidity in Dense Matter},
  author = {Page, Dany and Prakash, Madappa and Lattimer, James M. and Steiner, Andrew W.},
  journal = {Phys. Rev. Lett.},
  volume = {106},
  issue = {8},
  pages = {081101},
  numpages = {4},
  year = {2011},
  month = {Feb},
  publisher = {American Physical Society},
  doi = {10.1103/PhysRevLett.106.081101},
  url = {https://link.aps.org/doi/10.1103/PhysRevLett.106.081101}
}

@article{Shternin2011,
    author = {Shternin, Peter S. and Yakovlev, Dmitry G. and Heinke, Craig O. and Ho, Wynn C. G. and Patnaude, Daniel J.},
    title = {Cooling neutron star in the Cassiopeia A supernova remnant: evidence for superfluidity in the core},
    journal = {Monthly Notices of the Royal Astronomical Society: Letters},
    volume = {412},
    number = {1},
    pages = {L108-L112},
    year = {2011},
    month = {03},
    issn = {1745-3925},
    doi = {10.1111/j.1745-3933.2011.01015.x}
}

@article{Tabakin:1968zz,
    author = "Tabakin, Frank",
    title = "{Single Separable Potential with Attraction and Repulsion}",
    doi = "10.1103/PhysRev.174.1208",
    journal = "Phys. Rev.",
    volume = "174",
    pages = "1208--1212",
    year = "1968"
}

@article{Hoffberg:1970vqj,
    author = "Hoffberg, M. and Glassgold, A. E. and Richardson, R. W. and Ruderman, M.",
    title = "{Anisotropic Superfluidity in Neutron Star Matter}",
    doi = "10.1103/PhysRevLett.24.775",
    journal = "Phys. Rev. Lett.",
    volume = "24",
    number = "14",
    pages = "775",
    year = "1970"
}

@article{10.1143/PTP.44.905,
    author = {Tamagaki, Ryozo},
    title = {Superfluid State in Neutron Star Matter. I: Generalized Bogoliubov Transformation and Existence of 3P2 Gap at High Density},
    journal = {Progress of Theoretical Physics},
    volume = {44},
    number = {4},
    pages = {905-928},
    year = {1970},
    month = {10},
    issn = {0033-068X},
    doi = {10.1143/PTP.44.905},
    url = {https://doi.org/10.1143/PTP.44.905},
    eprint = {https://academic.oup.com/ptp/article-pdf/44/4/905/5386708/44-4-905.pdf},
}

@article{10.1143/PTP.46.114,
    author = {Takatsuka, Tatsuyuki and Tamagaki, Ryozo},
    title = {Superfluidity in Neutron Star Matter and Symmetric Nuclear Matter},
    journal = {Progress of Theoretical Physics Supplement},
    volume = {112},
    pages = {27-65},
    year = {1993},
    month = {03},
    issn = {0375-9687},
    doi = {10.1143/PTP.112.27},
    url = {https://doi.org/10.1143/PTP.112.27},
    eprint = {https://academic.oup.com/ptps/article-pdf/doi/10.1143/PTP.112.27/5205856/112-27.pdf},
}

@article{10.1143/PTP.47.1062,
    author = {Takatsuka, Tatsuyuki},
    title = {Superfluid State in Neutron Star Matter. III: Tensor Coupling Effect in 3P2 Energy Gap},
    journal = {Progress of Theoretical Physics},
    volume = {47},
    number = {3},
    pages = {1062-1064},
    year = {1972},
    month = {03},
    issn = {0033-068X},
    doi = {10.1143/PTP.47.1062},
    url = {https://doi.org/10.1143/PTP.47.1062},
    eprint = {https://academic.oup.com/ptp/article-pdf/47/3/1062/5443245/47-3-1062.pdf},
}

@article{Richardson:1972xn,
    author = "Richardson, R. W.",
    title = "{Ginzburg-landau theory of anisotropic superfluid neutron-star matter}",
    doi = "10.1103/PhysRevD.5.1883",
    journal = "Phys. Rev. D",
    volume = "5",
    pages = "1883--1896",
    year = "1972"
}

@article{Sauls:1978lna,
    author = "Sauls, J. and Serene, J.",
    title = "{$^{3}P_{2}$ pairing near the transition temperature in neutron-star matter}",
    doi = "10.1103/PhysRevD.17.1524",
    journal = "Phys. Rev. D",
    volume = "17",
    number = "6",
    pages = "1524--1528",
    year = "1978"
}

@article{Muzikar:1980as,
    author = "Muzikar, P. and Sauls, J. A. and Serene, J. W.",
    title = "{3P - 2 PAIRING IN NEUTRON STAR MATTER: MAGNETIC FIELD EFFECTS AND VORTICES}",
    doi = "10.1103/PhysRevD.21.1494",
    journal = "Phys. Rev. D",
    volume = "21",
    pages = "1494--1502",
    year = "1980"
}

@article{Sauls:1982ie,
    author = "Sauls, J. A. and Stein, D. L. and Serene, J. W.",
    title = "{MAGNETIC VORTICES IN A ROTATING p wave triplet NEUTRON SUPERFLUID}",
    doi = "10.1103/PhysRevD.25.967",
    journal = "Phys. Rev. D",
    volume = "25",
    pages = "967--975",
    year = "1982"
}

@article{Yasui:2018tcr,
    author = "Yasui, Shigehiro and Chatterjee, Chandrasekhar and Nitta, Muneto",
    title = "{Phase structure of neutron $^{3}P_{2}$ superfluids in strong magnetic fields in neutron stars}",
    eprint = "1810.04901",
    archivePrefix = "arXiv",
    primaryClass = "nucl-th",
    doi = "10.1103/PhysRevC.99.035213",
    journal = "Phys. Rev. C",
    volume = "99",
    number = "3",
    pages = "035213",
    year = "2019"
}

@article{Yasui:2019unp,
    author = "Yasui, Shigehiro and Chatterjee, Chandrasekhar and Kobayashi, Michikazu and Nitta, Muneto",
    title = "{Reexamining Ginzburg-Landau theory for neutron $^3P_2$ superfluidity in neutron stars}",
    eprint = "1904.11399",
    archivePrefix = "arXiv",
    primaryClass = "nucl-th",
    doi = "10.1103/PhysRevC.100.025204",
    journal = "Phys. Rev. C",
    volume = "100",
    number = "2",
    pages = "025204",
    year = "2019"
}

@article{Masuda:2016vak,
    author = "Masuda, Kota and Nitta, Muneto",
    title = "{Half-quantized non-Abelian vortices in neutron $^3P_2$ superfluids inside magnetars}",
    eprint = "1602.07050",
    archivePrefix = "arXiv",
    primaryClass = "nucl-th",
    doi = "10.1093/ptep/ptz138",
    journal = "PTEP",
    volume = "2020",
    number = "1",
    pages = "013D01",
    year = "2020"
}

@article{Yasui:2019vci,
    author = "Yasui, Shigehiro and Nitta, Muneto",
    title = "{Domain walls in neutron $^{3}P_{2}$ superfluids in neutron stars}",
    eprint = "1907.12843",
    archivePrefix = "arXiv",
    primaryClass = "nucl-th",
    doi = "10.1103/PhysRevC.101.015207",
    journal = "Phys. Rev. C",
    volume = "101",
    number = "1",
    pages = "015207",
    year = "2020"
}

@article{Song:2007ca,
    author = "Song, Jun Liang and Semenoff, Gordon W. and Zhou, Fei",
    title = "{Quantum fluctuation-induced uniaxial and biaxial spin nematics}",
    eprint = "cond-mat/0702052",
    archivePrefix = "arXiv",
    doi = "10.1103/PhysRevLett.98.160408",
    journal = "Phys. Rev. Lett.",
    volume = "98",
    pages = "160408",
    year = "2007"
}

@article{Uchino:2010pf,
    author = "Uchino, Shun and Kobayashi, Michikazu and Nitta, Muneto and Ueda, Masahito",
    title = "{Quasi-Nambu-Goldstone Modes in Bose-Einstein Condensates}",
    eprint = "1010.2864",
    archivePrefix = "arXiv",
    primaryClass = "cond-mat.quant-gas",
    doi = "10.1103/PhysRevLett.105.230406",
    journal = "Phys. Rev. Lett.",
    volume = "105",
    pages = "230406",
    year = "2010"
}

@article{Antonopoulou:2022rpq,
    author = "Antonopoulou, Danai and Haskell, Brynmor and Espinoza, Crist{\'o}bal M.",
    title = "{Pulsar glitches: observations and physical interpretation}",
    doi = "10.1088/1361-6633/ac9ced",
    journal = "Rept. Prog. Phys.",
    volume = "85",
    number = "12",
    pages = "126901",
    year = "2022"
}

@article{Zhou:2022cyp,
    author = {Zhou, Shiqi and G{\"u}gercino{\u{g}}lu, Erbil and Yuan, Jianping and Ge, Mingyu and Yu, Cong},
    title = "{Pulsar Glitches: A Review}",
    eprint = "2211.13885",
    archivePrefix = "arXiv",
    primaryClass = "astro-ph.HE",
    doi = "10.3390/universe8120641",
    journal = "Universe",
    volume = "8",
    number = "12",
    pages = "641",
    year = "2022"
}

@article{Sedrakian:2018ydt,
    author = "Sedrakian, Armen and Clark, John W.",
    title = "{Superfluidity in nuclear systems and neutron stars}",
    eprint = "1802.00017",
    archivePrefix = "arXiv",
    primaryClass = "nucl-th",
    doi = "10.1140/epja/i2019-12863-6",
    journal = "Eur. Phys. J. A",
    volume = "55",
    number = "9",
    pages = "167",
    year = "2019"
}

@article{Graber:2016imq,
    author = "Graber, Vanessa and Andersson, Nils and Hogg, Michael",
    title = "{Neutron Stars in the Laboratory}",
    eprint = "1610.06882",
    archivePrefix = "arXiv",
    primaryClass = "astro-ph.HE",
    doi = "10.1142/S0218271817300154",
    journal = "Int. J. Mod. Phys. D",
    volume = "26",
    number = "08",
    pages = "1730015",
    year = "2017"
}

@inbook{Antonelli_2023vpd,
author = {Marco Antonelli and Alessandro Montoli and Pierre M. Pizzochero},
title = {Insights Into the Physics of Neutron Star Interiors from Pulsar Glitches},
booktitle = {Astrophysics in the XXI Century with Compact Stars},
chapter = {Chapter 7},
pages = {219-281},
publisher = {WORLD SCIENTIFIC},
year = {2022},
doi = {10.1142/9789811220944{\_}0007},
URL = {https://www.worldscientific.com/doi/abs/10.1142/9789811220944_0007}
}

@article{Anderson:1975zze,
    author = "Anderson, P. W. and Itoh, N.",
    title = "{Pulsar glitches and restlessness as a hard superfluidity phenomenon}",
    doi = "10.1038/256025a0",
    journal = "Nature",
    volume = "256",
    pages = "25--27",
    year = "1975"
}

@article{Hinohara2024,
  title = {Triplet-odd pairing in finite nuclear systems: Even-even singly closed nuclei},
  author = {Hinohara, Nobuo and Oishi, Tomohiro and Yoshida, Kenichi},
  journal = {Phys. Rev. C},
  volume = {109},
  issue = {3},
  pages = {034302},
  numpages = {9},
  year = {2024},
  month = {Mar},
  publisher = {American Physical Society},
  doi = {10.1103/PhysRevC.109.034302},
  url = {https://link.aps.org/doi/10.1103/PhysRevC.109.034302}
}

@misc{Yoshimura2026,
      title={Superfluid Band Theory for the Rod Phase in the Magnetized Inner Crust Matter: Entrainment, Spin-orbit Coupling, Spin-triplet Pairing}, 
      author={Kenta Yoshimura and Kazuyuki Sekizawa},
      year={2026},
      eprint={2601.13636},
      archivePrefix={arXiv},
      primaryClass={nucl-th},
      url={https://arxiv.org/abs/2601.13636}, 
}

@article{Eto:2017rfr,
    author = "Eto, Minoru and Nitta, Muneto",
    title = "{Confinement of half-quantized vortices in coherently coupled Bose-Einstein condensates: Simulating quark confinement in a QCD-like theory}",
    eprint = "1702.04892",
    archivePrefix = "arXiv",
    primaryClass = "cond-mat.quant-gas",
    reportNumber = "YGHP-17-04",
    doi = "10.1103/PhysRevA.97.023613",
    journal = "Phys. Rev. A",
    volume = "97",
    number = "2",
    pages = "023613",
    year = "2018"
}

@article{Son:2001td,
    author = "Son, D. T. and Stephanov, Misha A.",
    title = "{Domain walls in two-component Bose-Einstein condensates}",
    eprint = "cond-mat/0103451",
    archivePrefix = "arXiv",
    doi = "10.1103/PhysRevA.65.063621",
    journal = "Phys. Rev. A",
    volume = "65",
    pages = "063621",
    year = "2002"
}

@article{Kasamatsu:2004tvg,
    author = "Kasamatsu, Kenichi and Tsubota, Makoto and Ueda, Masahito",
    title = "{Vortex molecules in coherently coupled two-component Bose-Einstein condensates}",
    eprint = "cond-mat/0406150",
    archivePrefix = "arXiv",
    doi = "10.1103/PhysRevLett.93.250406",
    journal = "Phys. Rev. Lett.",
    volume = "93",
    number = "25",
    pages = "250406",
    year = "2004"
}

@article{Kasamatsu:2005xiu,
    author = "Kasamatsu, Kenichi and Tsubota, Makoto and Ueda, Masahito",
    title = "{Vortices in Multicomponent Bose{\textendash}einstein Condensates}",
    eprint = "cond-mat/0505546",
    archivePrefix = "arXiv",
    doi = "10.1142/s0217979205029602",
    journal = "Int. J. Mod. Phys. B",
    volume = "19",
    number = "11",
    pages = "1835--1904",
    year = "2005"
}

@article{Cipriani:2013nya,
    author = "Cipriani, Mattia and Nitta, Muneto",
    title = "{Crossover between integer and fractional vortex lattices in coherently coupled two-component Bose-Einstein condensates}",
    eprint = "1303.2592",
    archivePrefix = "arXiv",
    primaryClass = "cond-mat.quant-gas",
    reportNumber = "IFUP-TH-2013-08",
    doi = "10.1103/PhysRevLett.111.170401",
    journal = "Phys. Rev. Lett.",
    volume = "111",
    pages = "170401",
    year = "2013"
}

@article{Eto:2019uhe,
    author = "Eto, Minoru and Ikeno, Kazuki and Nitta, Muneto",
    title = "{Collision dynamics and reactions of fractional vortex molecules in coherently coupled Bose-Einstein condensates}",
    eprint = "1912.09014",
    archivePrefix = "arXiv",
    primaryClass = "cond-mat.quant-gas",
    reportNumber = "YGHP-19-03",
    doi = "10.1103/PhysRevResearch.2.033373",
    journal = "Phys. Rev. Res.",
    volume = "2",
    number = "3",
    pages = "033373",
    year = "2020"
}

@article{Kobayashi:2018ezm,
    author = "Kobayashi, Michikazu and Eto, Minoru and Nitta, Muneto",
    title = "{Berezinskii-Kosterlitz-Thouless Transition of Two-Component Bose Mixtures with Intercomponent Josephson Coupling}",
    eprint = "1802.08763",
    archivePrefix = "arXiv",
    primaryClass = "cond-mat.stat-mech",
    doi = "10.1103/PhysRevLett.123.075303",
    journal = "Phys. Rev. Lett.",
    volume = "123",
    number = "7",
    pages = "075303",
    year = "2019"
}

@article{Tylutki:2016mgy,
    author = "Tylutki, Marek and Pitaevskii, Lev P. and Recati, Alessio and Stringari, Sandro",
    title = "{Confinement and precession of vortex pairs in coherently coupled Bose-Einstein condensates}",
    eprint = "1601.03695",
    archivePrefix = "arXiv",
    primaryClass = "cond-mat.quant-gas",
    doi = "10.1103/PhysRevA.93.043623",
    journal = "Phys. Rev. A",
    volume = "93",
    number = "4",
    pages = "043623",
    year = "2016"
}

@article{Eto:2012rc,
    author = "Eto, Minoru and Nitta, Muneto",
    title = "{Vortex trimer in three-component Bose-Einstein condensates}",
    eprint = "1201.0343",
    archivePrefix = "arXiv",
    primaryClass = "cond-mat.quant-gas",
    doi = "10.1103/PhysRevA.85.053645",
    journal = "Phys. Rev. A",
    volume = "85",
    pages = "053645",
    year = "2012"
}

@article{PhysRevA.95.023605,
  title = {Vortex dynamics in coherently coupled Bose-Einstein condensates},
  author = {Calderaro, Luca and Fetter, Alexander L. and Massignan, Pietro and Wittek, Peter},
  journal = {Phys. Rev. A},
  volume = {95},
  issue = {2},
  pages = {023605},
  numpages = {16},
  year = {2017},
  month = {Feb},
  publisher = {American Physical Society},
  doi = {10.1103/PhysRevA.95.023605},
  url = {https://link.aps.org/doi/10.1103/PhysRevA.95.023605}
}

@article{Eto:2013spa,
    author = "Eto, Minoru and Nitta, Muneto",
    title = "{Vortex graphs as N-omers and CP(N-1) Skyrmions in N-component Bose-Einstein condensates}",
    eprint = "1303.6048",
    archivePrefix = "arXiv",
    primaryClass = "cond-mat.quant-gas",
    doi = "10.1209/0295-5075/103/60006",
    journal = "EPL",
    volume = "103",
    number = "6",
    pages = "60006",
    year = "2013"
}

@article{Nitta:2010yf,
    author = "Nitta, Muneto and Eto, Minoru and Fujimori, Toshiaki and Ohashi, Keisuke",
    title = "{Baryonic Bound State of Vortices in Multicomponent Superconductors}",
    eprint = "1011.2552",
    archivePrefix = "arXiv",
    primaryClass = "cond-mat.supr-con",
    reportNumber = "RIKEN-MP-7, IFUP-TH-2010-41, KUNS-2311",
    doi = "10.1143/JPSJ.81.084711",
    journal = "J. Phys. Soc. Jap.",
    volume = "81",
    pages = "084711",
    year = "2012"
}

@article{Nitta:2013eaa,
    author = "Nitta, Muneto and Eto, Minoru and Cipriani, Mattia",
    title = "{Vortex molecules in Bose-Einstein condensates}",
    eprint = "1307.4312",
    archivePrefix = "arXiv",
    primaryClass = "cond-mat.quant-gas",
    doi = "10.1007/s10909-013-0925-3",
    journal = "J. Low Temp. Phys.",
    volume = "175",
    pages = "177--188",
    year = "2013"
}

@article{PhysRevA.94.023617,
  title = {Skyrmionic vortex lattices in coherently coupled three-component Bose-Einstein condensates},
  author = {Orlova, Natalia V. and Kuopanportti, Pekko and Milo\ifmmode \check{s}\else \v{s}\fi{}evi\ifmmode \acute{c}\else \'{c}\fi{}, Milorad V.},
  journal = {Phys. Rev. A},
  volume = {94},
  issue = {2},
  pages = {023617},
  numpages = {12},
  year = {2016},
  month = {Aug},
  publisher = {American Physical Society},
  doi = {10.1103/PhysRevA.94.023617},
  url = {https://link.aps.org/doi/10.1103/PhysRevA.94.023617}
}

@article{Dantas:2015fka,
    author = "Dantas, Davi S. and Lima, Aristeu R. P. and Chaves, A. and Almeida, C. A. S. and Farias, G. A. and Milo{\v{s}}evi{\'c}, M. V.",
    title = "{Bound vortex states and exotic lattices in multicomponent Bose-Einstein condensates: The role of vortex-vortex interaction}",
    eprint = "1504.03203",
    archivePrefix = "arXiv",
    primaryClass = "cond-mat.quant-gas",
    doi = "10.1103/PhysRevA.91.023630",
    journal = "Phys. Rev. A",
    volume = "91",
    pages = "023630",
    year = "2015"
}

@article{Garaud:2014laa,
    author = "Garaud, Julien and Babaev, Egor",
    title = "{Topological defects in mixtures of superconducting condensates with different charges}",
    eprint = "1403.3373",
    archivePrefix = "arXiv",
    primaryClass = "cond-mat.supr-con",
    doi = "10.1103/PhysRevB.89.214507",
    journal = "Phys. Rev. B",
    volume = "89",
    number = "21",
    pages = "214507",
    year = "2014"
}

@article{Chatterjee:2019jez,
   title={Chemical bonds of two vortex species with a generalized Josephson term and arbitrary charges},
   volume={2020},
   ISSN={1029-8479},
   url={http://dx.doi.org/10.1007/JHEP04(2020)109},
   DOI={10.1007/jhep04(2020)109},
   number={4},
   journal={Journal of High Energy Physics},
   publisher={Springer Science and Business Media LLC},
   author={Chatterjee, Chandrasekhar and Gudnason, Sven Bjarke and Nitta, Muneto},
   year={2020},
   month=Apr }

@article{Eto:2011wp,
    author = "Eto, Minoru and Kasamatsu, Kenichi and Nitta, Muneto and Takeuchi, Hiromitsu and Tsubota, Makoto",
    title = "{Interaction of half-quantized vortices in two-component Bose-Einstein condensates}",
    eprint = "1103.6144",
    archivePrefix = "arXiv",
    primaryClass = "cond-mat.quant-gas",
    doi = "10.1103/PhysRevA.83.063603",
    journal = "Phys. Rev. A",
    volume = "83",
    pages = "063603",
    year = "2011"
}

@article{
David_Edward_doi:10.1126/science.162.3861.1481,
author = {David H. Staelin  and Edward C. Reifenstein },
title = {Pulsating Radio Sources near the Crab Nebula},
journal = {Science},
volume = {162},
number = {3861},
pages = {1481-1483},
year = {1968},
doi = {10.1126/science.162.3861.1481},
URL = {https://www.science.org/doi/abs/10.1126/science.162.3861.1481},
eprint = {https://www.science.org/doi/pdf/10.1126/science.162.3861.1481}
}

@article{
A_Pulsar_Supernova_Association_Nature,
author = {LARGE M. and VAUGHAN A. and MILLS B.},
title = {OA Pulsar Supernova Association?},
journal = {Nature},
volume = {220},
pages = {340-341},
year = {1968},
doi = {10.1038/220340a0},
URL = {https://www.nature.com/articles/220340a0},
eprint = {https://www.nature.com/articles/220340a0}
}

@article{Alpar1984, 
   author = {M. A. Alpar, S. A. Langer, J. A. Sauls}, 
   journal = {Astrophys. J.},
   volume={282}, 
   pages={533},
   year={1984},
   title={Rapid postglitch spin-up of the superfluid core in pulsars},
   doi={10.1086/162232}
}

@ARTICLE{Mochizuki1995,
       author = {{Mochizuki}, Y. and {Izuyama}, T.},
        title = "{Self-trapping of Superfluid Vortices and the Origin of Pulsar Glitches}",
      journal = {\apj},
         year = 1995,
        month = feb,
       volume = {440},
        pages = {263},
          doi = {10.1086/175267},
       adsurl = {https://ui.adsabs.harvard.edu/abs/1995ApJ...440..263M}
}

@article{Mochizuki1997,
doi = {10.1086/304802},
url = {https://dx.doi.org/10.1086/304802},
year = {1997},
month = {nov},
publisher = {},
volume = {489},
number = {2},
pages = {848},
author = {Y. S. Mochizuki and K. Oyamatsu and T. Izuyama},
title = {Exotic Nuclear Rod Formation Induced by Superfluid Vortices in Neutron Star Crusts},
journal = {The Astrophysical Journal}
}

@article{Mochizuki1999,
doi = {10.1086/307509},
url = {https://dx.doi.org/10.1086/307509},
year = {1999},
month = {aug},
publisher = {},
volume = {521},
number = {1},
pages = {281},
author = {Y. Mochizuki and T. Izuyama and I. Tanihata},
title = {Dynamics of Exotic Nuclear Rod Formation for the Origin of Neutron Star Glitches},
journal = {The Astrophysical Journal}
}

@book{haensel2007neutron,
  title={Neutron Stars 1: Equation of State and Structure},
  author={Haensel, P. and Potekhin, A.Y. and Yakovlev, D.G.},
  isbn={9780387473017},
  lccn={2006923828},
  series={Astrophysics and Space Science Library},
  url={https://www.springer.com/gp/book/9780387335438},
  year={2007},
  publisher={Springer New York}
}

@article{10.1093/mnras/sts108,
    author = {Warszawski, L. and Melatos, A.},
    title = "{Knock-on processes in superfluid vortex avalanches and pulsar glitch statistics}",
    journal = {Mon. Not. R. Astrn. Soc.},
    volume = {428},
    number = {3},
    pages = {1911-1926},
    year = {2012},
    issn = {0035-8711},
    doi = {10.1093/mnras/sts108},
    url = {https://doi.org/10.1093/mnras/sts108}
}

@article{PhysRevLett.109.241103,
  title = {Pulsar Glitches: The Crust is not Enough},
  author = {Andersson, N. and Glampedakis, K. and Ho, W. C. G. and Espinoza, C. M.},
  journal = {Phys. Rev. Lett.},
  volume = {109},
  issue = {24},
  pages = {241103},
  numpages = {5},
  year = {2012},
  publisher = {American Physical Society},
  doi = {10.1103/PhysRevLett.109.241103},
  url = {https://link.aps.org/doi/10.1103/PhysRevLett.109.241103}
}

@article{PhysRevLett.110.011101,
  title = {Crustal Entrainment and Pulsar Glitches},
  author = {Chamel, N.},
  journal = {Phys. Rev. Lett.},
  volume = {110},
  issue = {1},
  pages = {011101},
  numpages = {5},
  year = {2013},
  publisher = {American Physical Society},
  doi = {10.1103/PhysRevLett.110.011101},
  url = {https://link.aps.org/doi/10.1103/PhysRevLett.110.011101}
}

@article{PhysRevC.90.015803,
  title = {Pulsar glitches: The crust may be enough},
  author = {Piekarewicz, J. and Fattoyev, F. J. and Horowitz, C. J.},
  journal = {Phys. Rev. C},
  volume = {90},
  issue = {1},
  pages = {015803},
  numpages = {11},
  year = {2014},
  publisher = {American Physical Society},
  doi = {10.1103/PhysRevC.90.015803},
  url = {https://link.aps.org/doi/10.1103/PhysRevC.90.015803}
}

@article{Babaev:2001hv,
    author = "Babaev, Egor",
    title = "{Vortices carrying an arbitrary fraction of magnetic flux quantum in two gap superconductors}",
    eprint = "cond-mat/0111192",
    archivePrefix = "arXiv",
    doi = "10.1103/PhysRevLett.89.067001",
    journal = "Phys. Rev. Lett.",
    volume = "89",
    pages = "067001",
    year = "2002"
}

@article{Babaev:2004rm,
    author = "Babaev, Egor and Sudbo, Asle and Ashcroft, N. W.",
    title = "{A Superconductor to superfluid phase transition in liquid metallic hydrogen}",
    eprint = "cond-mat/0410408",
    archivePrefix = "arXiv",
    doi = "10.1038/nature02910",
    journal = "Nature",
    volume = "431",
    pages = "666--668",
    year = "2004"
}

@article{Babaev:2002ck,
    author = "Babaev, Egor",
    title = "{Phase diagram of a planar two band superconductor: Condensation of vortices with fractional flux quantum and existence of a nonsuperconducting superfluid state in this system}",
    eprint = "cond-mat/0201547",
    archivePrefix = "arXiv",
    doi = "10.1016/j.nuclphysb.2004.02.021",
    journal = "Nucl. Phys. B",
    volume = "686",
    pages = "397",
    year = "2004"
}

@article{Goryo:2007qlm,
    author = "Goryo, Jun and Soma, Singo and Matsukawa, Hiroshi",
    title = "{Deconfinement of Vortices with Continuously Variable Fractions of the Unit Flux Quanta in Two-Gap Superconductors}",
    eprint = "cond-mat/0608015",
    archivePrefix = "arXiv",
    doi = "10.1209/0295-5075/80/17002",
    journal = "EPL",
    volume = "80",
    pages = "17002",
    year = "2007"
}

@article{Garaud:2015,
author = "Garaud, Julien and Babaev, Egor",
year = "2015",
title = "
Properties of skyrmions and multi-quanta vortices in chiral p-wave superconductors",
journal = "Scientific Reports",
pages = "17540",
volume =  "5",
doi =  "10.1038/srep17540",
}

@article{PhysRevB.86.060514,
  title = {Skyrmionic state and stable half-quantum vortices in chiral $p$-wave superconductors},
  author = {Garaud, Julien and Babaev, Egor},
  journal = {Phys. Rev. B},
  volume = {86},
  issue = {6},
  pages = {060514(R)},
  numpages = {5},
  year = {2012},
  month = {Aug},
  publisher = {American Physical Society},
  doi = {10.1103/PhysRevB.86.060514},
  url = {https://link.aps.org/doi/10.1103/PhysRevB.86.060514}
}

@article{NamSekizawa2025,
  author  = {Nam, Y. B. and Sekizawa, K.},
  title   = {Vortex Creep Heating in Neutron Star Cooling: 
             New Insights into Thermal Evolution of Heavy Neutron Stars},
  journal = {arXiv preprint},
  year    = {2025},
  eprint  = {2510.24167},
  archivePrefix = {arXiv}
}

@article{NamSekizawa2026,
  title = {Data-driven exploration of the neutron $^{3}P_{2}$ pairing gap using Cassiopeia A neutron star observational data: Direct ${\ensuremath{\chi}}^{2}$ minimization},
  author = {Nam, Yoonhak and Sekizawa, Kazuyuki},
  journal = {Phys. Rev. C},
  volume = {113},
  issue = {4},
  pages = {045807},
  numpages = {18},
  year = {2026},
  month = {Apr},
  publisher = {American Physical Society},
  doi = {10.1103/ssg9-xl98},
  url = {https://link.aps.org/doi/10.1103/ssg9-xl98}
}

@article{Negreiros2013,
title = {Impact of rotation-driven particle repopulation on the thermal evolution of pulsars},
journal = {Physics Letters B},
volume = {718},
number = {4},
pages = {1176-1180},
year = {2013},
issn = {0370-2693},
doi = {https://doi.org/10.1016/j.physletb.2012.12.046},
url = {https://www.sciencedirect.com/science/article/pii/S0370269312012993},
author = {Rodrigo Negreiros and Stefan Schramm and Fridolin Weber}
}

@misc{Sedrakian:2026jkj,
      title={Spin effects in superfluidity, neutron matter and neutron stars}, 
      author={Armen Sedrakian and Peter B. Rau},
      year={2026},
      eprint={2604.02782},
      archivePrefix={arXiv},
      primaryClass={astro-ph.HE},
      url={https://arxiv.org/abs/2604.02782}, 
}

@article{Sedrakian:2024dgk,
    author = "Sedrakian, Armen and Rau, Peter B.",
    title = "{Josephson currents in neutron stars}",
    eprint = "2407.13686",
    archivePrefix = "arXiv",
    primaryClass = "astro-ph.HE",
    reportNumber = "INT-PUB-24-031",
    doi = "10.1103/PhysRevD.111.023044",
    journal = "Phys. Rev. D",
    volume = "111",
    number = "2",
    pages = "023044",
    year = "2025"
}

@article{JOSEPHSON1962251,
title = {Possible new effects in superconductive tunnelling},
journal = {Physics Letters},
volume = {1},
number = {7},
pages = {251-253},
year = {1962},
issn = {0031-9163},
doi = {https://doi.org/10.1016/0031-9163(62)91369-0},
url = {https://www.sciencedirect.com/science/article/pii/0031916362913690},
author = {B.D. Josephson}
}

@article{Guo:2022xhc,
    author = "Guo, Yixin and Tajima, Hiroyuki",
    title = "{Competition between pairing and tripling in one-dimensional fermions with coexistent s- and p-wave interactions}",
    eprint = "2210.07042",
    archivePrefix = "arXiv",
    primaryClass = "cond-mat.quant-gas",
    reportNumber = "RIKEN-iTHEMS-Report-22",
    doi = "10.1103/PhysRevB.107.024511",
    journal = "Phys. Rev. B",
    volume = "107",
    number = "2",
    pages = "024511",
    year = "2023"
}

@Inbook{sauls_nato,
author="Sauls, J. A.",
editor="{\"O}gelman, H.
and van den Heuvel, E. P. J.",
title="Superfluidity in the Interiors of Neutron Stars",
bookTitle="Timing Neutron Stars",
year="1989",
publisher="Springer Netherlands",
address="Dordrecht",
pages="457--490",
doi="10.1007/978-94-009-2273-0_43",
}

@article{bedaquePRC14,
  title = {Neutrino emissivity from Goldstone boson decay in magnetized neutron matter},
  author = {Bedaque, Paulo and Sen, Srimoyee},
  journal = {Phys. Rev. C},
  volume = {89},
  issue = {3},
  pages = {035808},
  numpages = {6},
  year = {2014},
  month = {Mar},
  publisher = {American Physical Society},
  doi = {10.1103/PhysRevC.89.035808},
}

@article{Leinson:2011wf,
      author         = "Leinson, L. B.",
      title          = "{New eigen-mode of spin oscillations in the triplet
                        superfluid condensate in neutron stars}",
      journal        = "Phys. Lett.",
      volume         = "B702",
      year           = "2011",
      pages          = "422-428",
      doi            = "10.1016/j.physletb.2011.07.025",
      eprint         = "1107.4025",
      archivePrefix  = "arXiv",
      primaryClass   = "nucl-th",
      SLACcitation   = "%%CITATION = ARXIV:1107.4025;%%"
}

@article{Yasui:2019pgb,
    author = "Yasui, Shigehiro and Chatterjee, Chandrasekhar and Nitta, Muneto",
    title = "{Topological defects at the boundary of neutron $^{3}P_{2}$ superfluids in neutron stars}",
    eprint = "1905.13666",
    archivePrefix = "arXiv",
    primaryClass = "nucl-th",
    doi = "10.1103/PhysRevC.101.025204",
    journal = "Phys. Rev. C",
    volume = "101",
    number = "2",
    pages = "025204",
    year = "2020"
}

@article{Baldo:1998ca,
      author         = "Baldo, M. and Elgaroey, O. and Engvik, L. and
                        Hjorth-Jensen, M. and Schulze, H. J.",
      title          = "{Triplet P-3(2) to F-3(2) pairing in neutron matter with
                        modern nucleon-nucleon potentials}",
      journal        = "Phys. Rev.",
      volume         = "C58",
      year           = "1998",
      pages          = "1921-1928",
      doi            = "10.1103/PhysRevC.58.1921",
      eprint         = "nucl-th/9806097",
      archivePrefix  = "arXiv",
      primaryClass   = "nucl-th",
      SLACcitation   = "%%CITATION = NUCL-TH/9806097;%%"
}

@article{Takatsuka:1992ga,
      author         = "Takatsuka, Tatsuyuki and Tamagaki, Ryozo",
      title          = "{Superfluidity in neutron star matter and symmetric
                        nuclear matter}",
      journal        = "Prog. Theor. Phys. Suppl.",
      volume         = "112",
      year           = "1993",
      pages          = "27-66",
      doi            = "10.1143/PTPS.112.27",
      reportNumber   = "KUNS-1149",
      SLACcitation   = "%%CITATION = PTPSA,112,27;%%"
}

@article{Khodel:2000qw,
      author         = "Khodel, V. V. and Khodel, V. A. and Clark, John Walter",
      title          = "{Triplet pairing in neutron matter}",
      journal        = "Nucl. Phys.",
      volume         = "A679",
      year           = "2001",
      pages          = "827-867",
      doi            = "10.1016/S0375-9474(00)00351-1",
      eprint         = "nucl-th/0001006",
      archivePrefix  = "arXiv",
      primaryClass   = "nucl-th",
      SLACcitation   = "%%CITATION = NUCL-TH/0001006;%%"
}

@article{Bogner:2009bt,
      author         = "Bogner, S. K. and Furnstahl, R. J. and Schwenk, A.",
      title          = "{From low-momentum interactions to nuclear structure}",
      journal        = "Prog. Part. Nucl. Phys.",
      volume         = "65",
      year           = "2010",
      pages          = "94-147",
      doi            = "10.1016/j.ppnp.2010.03.001",
      eprint         = "0912.3688",
      archivePrefix  = "arXiv",
      primaryClass   = "nucl-th",
      SLACcitation   = "%%CITATION = ARXIV:0912.3688;%%"
}

@article{Maurizio:2014qsa,
      author         = "Maurizio, Stefano and Holt, Jeremy W. and Finelli, Paolo",
      title          = "{Nuclear pairing from microscopic forces: singlet
                        channels and higher-partial waves}",
      journal        = "Phys. Rev.",
      volume         = "C90",
      year           = "2014",
      number         = "4",
      pages          = "044003",
      doi            = "10.1103/PhysRevC.90.044003",
      eprint         = "1408.6281",
      archivePrefix  = "arXiv",
      primaryClass   = "nucl-th",
      SLACcitation   = "%%CITATION = ARXIV:1408.6281;%%"
}

@article{Srinivas:2016kir,
      author         = "Srinivas, Sarath and Ramanan, S.",
      title          = "{Triplet Pairing in pure neutron matter}",
      journal        = "Phys. Rev.",
      volume         = "C94",
      year           = "2016",
      number         = "6",
      pages          = "064303",
      doi            = "10.1103/PhysRevC.94.064303",
      eprint         = "1606.09053",
      archivePrefix  = "arXiv",
      primaryClass   = "nucl-th",
      SLACcitation   = "%%CITATION = ARXIV:1606.09053;%%"
}

@article{Amundsen:1984qc,
      author         = "Amundsen, L. and Ostgaard, E.",
      title          = "{Superfluidity of neutron matter (II). triplet pairing}",
      journal        = "Nucl. Phys.",
      volume         = "A442",
      year           = "1985",
      pages          = "163-188",
      doi            = "10.1016/0375-9474(85)90140-X,
                        10.1016/S0375-9474(85)80012-9",
      reportNumber   = "TRONDHEIM-84-16",
      SLACcitation   = "%%CITATION = NUPHA,A442,163;%%"
}

@article{Baldo:1992kzz,
      author         = "Baldo, M. and Cugnon, J. and Lejeune, A. and Lombardo,
                        U.",
      title          = "{Proton and neutron superfluidity in neutron star
                        matter}",
      journal        = "Nucl. Phys.",
      volume         = "A536",
      year           = "1992",
      pages          = "349-365",
      doi            = "10.1016/0375-9474(92)90387-Y",
      SLACcitation   = "%%CITATION = NUPHA,A536,349;%%"
}

@article{Elgaroy:1996hp,
      author         = "Elgaroy, O. and Engvik, L. and Hjorth-Jensen, M. and
                        Osnes, E.",
      title          = "{Triplet pairing of neutrons in beta stable neutron star
                        matter}",
      journal        = "Nucl. Phys.",
      volume         = "A607",
      year           = "1996",
      pages          = "425-441",
      doi            = "10.1016/0375-9474(96)00217-5",
      eprint         = "nucl-th/9604032",
      archivePrefix  = "arXiv",
      primaryClass   = "nucl-th",
      reportNumber   = "ECT-96-003",
      SLACcitation   = "%%CITATION = NUCL-TH/9604032;%%"
}

@article{Khodel:1998hn,
      author         = "Khodel, V. A. and Khodel, V. V. and Clark, John Walter",
      title          = "{Universalities of triplet pairing in neutron matter}",
      journal        = "Phys. Rev. Lett.",
      volume         = "81",
      year           = "1998",
      pages          = "3828-3831",
      doi            = "10.1103/PhysRevLett.81.3828",
      eprint         = "nucl-th/9807034",
      archivePrefix  = "arXiv",
      primaryClass   = "nucl-th",
      SLACcitation   = "%%CITATION = NUCL-TH/9807034;%%"
}

@article{Zverev:2003ak,
      author         = "Zverev, M. V. and Clark, John Walter and Khodel, V. A.",
      title          = "{$^3P_2$-$^3F_2$ pairing in dense neutron matter: The
                        Spectrum of solutions}",
      journal        = "Nucl. Phys.",
      volume         = "A720",
      year           = "2003",
      pages          = "20-42",
      doi            = "10.1016/S0375-9474(03)00653-5",
      eprint         = "nucl-th/0301028",
      archivePrefix  = "arXiv",
      primaryClass   = "nucl-th",
      SLACcitation   = "%%CITATION = NUCL-TH/0301028;%%"
}

@article{Bedaque:2003wj,
      author         = "Bedaque, Paulo F. and Rupak, Gautam and Savage, Martin
                        J.",
      title          = "{Goldstone bosons in the $^3P_2$ superfluid phase of
                        neutron matter and neutrino emission}",
      journal        = "Phys. Rev.",
      volume         = "C68",
      year           = "2003",
      pages          = "065802",
      doi            = "10.1103/PhysRevC.68.065802",
      eprint         = "nucl-th/0305032",
      archivePrefix  = "arXiv",
      primaryClass   = "nucl-th",
      reportNumber   = "NT-UW-03-010, LBNL-52649",
      SLACcitation   = "%%CITATION = NUCL-TH/0305032;%%"
}

@article{Bedaque:2012bs,
      author         = "Bedaque, Paulo F. and Nicholson, Amy N.",
      title          = "{Low lying modes of triplet-condensed neutron matter and
                        their effective theory}",
      journal        = "Phys. Rev.",
      volume         = "C87",
      year           = "2013",
      number         = "5",
      pages          = "055807",
      doi            = "10.1103/PhysRevC.89.029902, 10.1103/PhysRevC.87.055807",
      note           = "[Erratum: Phys. Rev.C89,no.2,029902(2014)]",
      eprint         = "1212.1122",
      archivePrefix  = "arXiv",
      primaryClass   = "nucl-th",
      reportNumber   = "UM-DOE-ER-40762-527",
      SLACcitation   = "%%CITATION = ARXIV:1212.1122;%%"
}

@article{Bedaque:2013fja,
      author         = "Bedaque, Paulo F. and Reddy, Sanjay",
      title          = "{Goldstone modes in the neutron star core}",
      journal        = "Phys. Lett.",
      volume         = "B735",
      year           = "2014",
      pages          = "340-343",
      doi            = "10.1016/j.physletb.2014.06.033",
      eprint         = "1307.8183",
      archivePrefix  = "arXiv",
      primaryClass   = "nucl-th",
      reportNumber   = "INT-PUB-13-031",
      SLACcitation   = "%%CITATION = ARXIV:1307.8183;%%"
}

@article{Bedaque:2014zta,
      author         = "Bedaque, Paulo F. and Nicholson, Amy N. and Sen,
                        Srimoyee",
      title          = "{Massive and massless modes of the triplet phase of
                        neutron matter}",
      journal        = "Phys. Rev.",
      volume         = "C92",
      year           = "2015",
      number         = "3",
      pages          = "035809",
      doi            = "10.1103/PhysRevC.92.035809",
      eprint         = "1408.5145",
      archivePrefix  = "arXiv",
      primaryClass   = "nucl-th",
      reportNumber   = "UM-DOE-ER-40762-527",
      SLACcitation   = "%%CITATION = ARXIV:1408.5145;%%"
}

@article{Leinson:2009nu,
      author         = "Leinson, L. B.",
      title          = "{Neutrino emission from triplet pairing of neutrons in
                        neutron stars}",
      journal        = "Phys. Rev.",
      volume         = "C81",
      year           = "2010",
      pages          = "025501",
      doi            = "10.1103/PhysRevC.81.025501",
      eprint         = "0912.2164",
      archivePrefix  = "arXiv",
      primaryClass   = "astro-ph.SR",
      SLACcitation   = "%%CITATION = ARXIV:0912.2164;%%"
}

@article{Leinson:2010yf,
      author         = "Leinson, L. B.",
      title          = "{Neutrino emission from spin waves in neutron
                        spin-triplet superfluid}",
      journal        = "Phys. Lett.",
      volume         = "B689",
      year           = "2010",
      pages          = "60-65",
      doi            = "10.1016/j.physletb.2010.04.046",
      eprint         = "1001.2617",
      archivePrefix  = "arXiv",
      primaryClass   = "astro-ph.SR",
      SLACcitation   = "%%CITATION = ARXIV:1001.2617;%%"
}

@article{Leinson:2010pk,
      author         = "Leinson, L. B.",
      title          = "{Superfluid phases of triplet pairing and neutrino
                        emission from neutron stars}",
      journal        = "Phys. Rev.",
      volume       = "C82",
      year           =  "2010",
      pages        =  "065503",
      doi             = "10.1103/PhysRevC.82.065503",
      eprint         = "1012.5387",
      archivePrefix  = "arXiv",
      primaryClass   = "hep-ph",
      SLACcitation   = "%%CITATION = ARXIV:1012.5387;%%"
}

@article{Leinson:2010ru,
      author         = "Leinson, L. B.",
      title          = "{Zero sound in triplet-correlated superfluid neutron
                        matter}",
      journal        = "Phys. Rev.",
      volume         = "C83",
      year           = "2011",
      pages          = "055803",
      doi            = "10.1103/PhysRevC.83.055803",
      eprint         = "1007.2803",
      archivePrefix  = "arXiv",
      primaryClass   = "hep-ph",
      SLACcitation   = "%%CITATION = ARXIV:1007.2803;%%"
}

@article{Leinson:2011jr,
      author         = "Leinson, L. B.",
      title          = "{Neutrino emissivity of $^{3}P_{2}$-$^{3}F_{2}$
                        superfluid cores in neutron stars}",
      journal        = "Phys. Rev.",
      volume         = "C84",
      year           = "2011",
      pages          = "045501",
      doi            = "10.1103/PhysRevC.84.045501",
      eprint         = "1110.2145",
      archivePrefix  = "arXiv",
      primaryClass   = "nucl-th",
      SLACcitation   = "%%CITATION = ARXIV:1110.2145;%%"
}

@article{Leinson:2012pn,
      author         = "Leinson, L. B.",
      title          = "{Collective modes of the order parameter in a triplet
                        superfluid neutron liquid}",
      journal        = "Phys. Rev.",
      volume         = "C85",
      year           = "2012",
      pages          = "065502",
      doi            = "10.1103/PhysRevC.85.065502",
      eprint         = "1206.3648",
      archivePrefix  = "arXiv",
      primaryClass   = "nucl-th",
      SLACcitation   = "%%CITATION = ARXIV:1206.3648;%%"
}

@article{Leinson:2013si,
      author         = "Leinson, L. B.",
      title          = "{Neutrino emissivity of anisotropic neutron superfluids}",
      journal        = "Phys. Rev.",
      volume         = "C87",
      year           = "2013",
      number         = "2",
      pages          = "025501",
      doi            = "10.1103/PhysRevC.87.025501",
      eprint         = "1301.5439",
      archivePrefix  = "arXiv",
      primaryClass   = "nucl-th",
      SLACcitation   = "%%CITATION = ARXIV:1301.5439;%%"
}

\end{document}